\documentclass[5p,twocolumn,authoryear,times]{elsarticle}

\usepackage{lineno,hyperref}
\modulolinenumbers[5]
\hypersetup{colorlinks,citecolor=blue,linkcolor=blue,urlcolor=blue}

\journal{Astronomy \& Computing}

\usepackage{graphicx}	
\usepackage{amsmath}	
\usepackage{amssymb}	

\usepackage{natbib} 
\usepackage[usenames,dvipsnames]{color}
\usepackage{verbatim}
\usepackage{multirow}
\usepackage{booktabs}
\usepackage{subfigure}
\usepackage{colortbl}

\usepackage[T1]{fontenc}
\usepackage[tracking=smallcaps]{microtype}
\SetTracking[no ligatures = f]{encoding = *,shape = sc}{0}

\usepackage{listings}
\usepackage[font=footnotesize,labelfont=bf]{caption}
\usepackage{footnote}
\makesavenoteenv{tabular}
\makesavenoteenv{table}
\usepackage{supertabular}
\usepackage{threeparttable}
\usepackage{longtable}
\usepackage{rotating}
\usepackage{lscape}

\usepackage{lineno}

\usepackage{bold-extra}
\usepackage{slantsc}
\usepackage{times}

\definecolor{verbgray}{gray}{0.9}
\definecolor{shadecolor}{rgb}{.9, .9, .9}
\definecolor{lightred}{rgb}{0.96,0.81,0.81}
\definecolor{verylightceleste}{rgb}{0.91, 0.93, 0.97}
\definecolor{lightceleste}{rgb}{0.75, 0.81, 0.98}
\definecolor{celeste}{rgb}{0.45, 0.58, 0.82}
\definecolor{white}{rgb}{1., 1., 1.}
\definecolor{darkblue}{rgb}{0.10, 0.35, 0.99}
\definecolor{darkblue2}{rgb}{0.10, 0.24, 0.59}
\definecolor{verydarkblue}{rgb}{0.05, 0.10, 0.24}
\definecolor{black}{rgb}{0., 0., 0.}
\definecolor{lightviolet}{rgb}{0.95, 0.89, 0.96}
\definecolor{violet}{rgb}{0.7, 0.28, 0.70}
\definecolor{darkviolet}{rgb}{0.42, 0.24, 0.41}
\definecolor{green}{rgb}{0.18, 0.41, 0.02}
\definecolor{lightgreen}{rgb}{0.62, 0.75, 0.43}
\definecolor{darkgreen}{rgb}{0.27, 0.34, 0.16}
\definecolor{verylightgreen}{rgb}{0.298039,0.952941,0.345098}
\definecolor{verylightgreen2}{rgb}{0.745098,1,0.654902}
\definecolor{yellow}{rgb}{1, 1, 0}
\definecolor{darkyellow2}{rgb}{1, 0.890196,0.427451}
\definecolor{darkyellow}{rgb}{0.87, 0.87, 0.03}
\definecolor{darkyellow3}{rgb}{1,0.933333,0.654902}
\definecolor{gray}{rgb}{0.14, 0.14, 0.15}
\definecolor{lightgray}{rgb}{0.70, 0.72, 0.74}
\definecolor{silver}{rgb}{0.91, 0.91, 0.91}
\definecolor{Blueb}{cmyk}{0.2,0,0,0}
\definecolor{lightorange}{rgb}{1,0.886275,0.619608}

\newcommand{\caesar}{\emph{caesar}}
\newcommand{\scorpio}{\textsc{Scorpio}}
\newcommand{\toolname}{\emph{caesar-mrcnn}}

\usepackage{xcolor}
\definecolor{light-gray}{gray}{0.95}

\lstnewenvironment{codebkg}{%
  \lstset{
    escapechar=\&,
    basicstyle=\ttfamily\scriptsize,
    backgroundcolor=\color{light-gray},
    frame=single,
    breaklines=true,    
    framerule=0pt,
    columns=fullflexible
  }
 }
 {}

\begin{document}
\begin{frontmatter}

\title{Astronomical source detection in radio continuum maps with deep neural networks}

\author[1]{S. Riggi\corref{cor}}
\ead{simone.riggi@inaf.it}
\author[2,1]{D. Magro}%
\author[3,1]{R. Sortino}%
\author[2]{A. De Marco}%
\author[1]{C. Bordiu}%
\author[4]{T. Cecconello}%
\author[5]{A. M. Hopkins}%
\author[6]{J. Marvil}%
\author[1]{G. Umana}%
\author[1]{E. Sciacca}%
\author[7]{F. Vitello}%
\author[1]{F. Bufano}%
\author[1]{A. Ingallinera}%
\author[8]{G. Fiameni}%
\author[3]{C. Spampinato}%
\author[2,9]{K. Zarb Adami}%

\cortext[cor]{Corresponding author}
\address[1]{INAF-Osservatorio Astrofisico di Catania, Via Santa Sofia 78, 95123, Catania, Italy}%
\address[2]{Institute of Space Sciences and Astronomy, University of Malta, Msida MSD2080, Malta}
\address[3]{Department of Electrical, Electronic and Computer Engineering, University of Catania, Viale Andrea Doria 6, 95125, Catania, Italy}
\address[4]{Complex Systems and Artificial Intelligence research center, University of Milano-Bicocca, Viale Sarca 336, 20126, Milano, Italy}
\address[5]{Australian Astronomical Optics, Macquarie University, 105 Delhi Rd, North Ryde, NSW 2113, Australia}%
\address[6]{National Radio Astronomy Observatory, P.O. Box O, Socorro, NM 87801, USA}
\address[7]{INAF-Istituto. di Radioastronomia, Via Gobetti 101 40127 Bologna, Italy}
\address[8]{NVIDIA AI Technology Centre, Italy}
\address[9]{Department of Astrophysics, University of Oxford, Oxford OX1 2JD, United Kingdom}

%

\begin{abstract}
Source  finding  is  one  of  the  most  challenging  tasks in upcoming radio continuum surveys with SKA precursors, such as the Evolutionary Map of the Universe (EMU) survey of the Australian SKA Pathfinder (ASKAP) telescope. The resolution, sensitivity, and sky coverage of such surveys is unprecedented, requiring new features and improvements to be made in existing source finders. Among them, reducing the false detection rate, particularly in the Galactic plane, and the ability to associate multiple disjoint islands into physical objects.
To bridge this gap, we developed a new source finder, based on the Mask R-CNN object detection framework, capable of both detecting and classifying compact, extended, spurious, and poorly imaged sources in radio continuum images. The model was trained using ASKAP EMU data, observed during the Early Science and pilot survey phase, and previous radio survey data, taken with the VLA and ATCA telescopes.
On the test sample, the final model achieves an overall detection completeness above 85\%, a reliability of $\sim$65\%, and a classification precision/recall above 90\%. Results obtained for all source classes are reported and discussed.
\end{abstract}

\begin{keyword}
radio continuum: general \sep techniques: image processing \sep source finding \sep SKA precursors \sep Neural networks \sep Instance segmentation
\end{keyword}

\end{frontmatter}

\section{Introduction}
\label{sec:intro}
Source finding is one of the most challenging tasks in upcoming radio continuum surveys, such as the Evolutionary Map of the Universe (EMU) \citep{Norris2011} planned at the Australian SKA Pathfinder (ASKAP) telescope \citep{Johnston2008,ASKAPSystemDesign}. The enhanced sensitivity, angular resolution and field of view will enable the detection of millions of sources in EMU. Such cataloguing process requires a level of automation and accuracy in knowledge extraction never reached before in source finder algorithms. 

Although existing finders used in the radio community
have been recently upgraded in this direction (e.g. see \citealt{Hancock2018,Riggi2019,Carbone2018}), and novel solutions have been developed (e.g. see \citealt{Robotham2018,Hale2019,Lukas2019}), many algorithmic aspects, critical for EMU but also for future SKA surveys, remain to be tackled, particularly for observations carried out in the Galactic plane.\\
Identification of spurious sources is certainly a priority, particularly for observations with a significant diffuse background or very extended sources, where the false detection rate obtained by standard source finders can exceed the 20\% level \citep{Riggi2021}. 
Spurious detections in island\footnote{In source finding terminology, an island (or blob) denotes a group of 4-connected pixels in the image under analysis with brightness above a "merge" (or "aggregation") threshold (usually chosen in the range 2.5-3.0 $\sigma$ with respect to the image background), and around a "seed" pixel with brightness above a detection threshold (usually 5$\sigma$).} catalogues (and consequently also in the corresponding fitted component catalogues) are due to the background noise and artefacts introduced in the imaging process. Among them, sidelobes around bright sources, dominate at high S/N levels, e.g. well above the standard 5$\sigma$ detection threshold. Spurious detections in component catalogues are partly due to the presence of spurious islands but, mostly, to the island over-deblending of existing finders, particularly when estimating components of extended or non-gaussian islands. These spurious detections are only rarely automatically rejected in large area surveys (for example using ad-hoc selection cuts), e.g. the most widely-used approach is identifying them by visual inspection.\\ 
Source identification from multiple non-contiguous islands and classification into known classes of astrophysical objects is another poorly covered task in traditional source finders. This is particularly relevant when searching for Galactic objects in Galactic plane surveys. In these studies, the extragalactic objects constitute the most numerous background sources ($\sim90\%$ of the catalogued sources). Although the majority of them have a single-island morphology, radio galaxies
with an extended morphology (e.g. including multiple islands associated to their physical core, lobe, or jet components) can easily exceed the number of Galactic objects previously known in the considered map. For instance, in the ASKAP \textsc{Scorpio} survey \citep{Riggi2021}, the number of islands associated to radio galaxies was found to be a factor $\sim$3 larger than those associated to known or candidate Galactic objects previously reported in the literature. Identification and removal of this kind of sources would therefore ease the search of unclassified Galactic objects.\\Machine learning was already proven to be a valuable tool for tackling most of the aforementioned tasks. For example, \citep{Lukic2018,Lukic2020,Wu2019} employed deep Convolutional Neural Networks (CNNs) for detecting and classifying radio galaxies in extragalactic fields. New source finders, based on deep networks, were also recently implemented and made available to the radio community. \emph{ConvoSource} \citep{Lukic2020}, for instance, is a CNN-based tool for semantic segmentation of radio sources. It was trained on a dataset composed of simulated compact and extended star-forming galaxies (SFGs) and Active Galactic Nuclei (AGN) (both steep- and flat-spectrum populations), as modelled in the SKA Data Challenge I (SDC1) \citep{Bonaldi2020}. Best performances (precision=0.73, recall=0.83, F1-score=0.78, all classes, SNR>5) were obtained on SKA Band 1 simulated maps with high integration times (1000 h).\\
\emph{ClaRAN} \citep{Wu2019}, a Faster R-CNN \citep{Ren2017} based model, detects and classifies radio galaxies of different morphological classes, with overall Mean Average Precision (mAP) ranging from 0.77 to 0.84, depending on the data pre-processing used. It exploits both real radio and infrared input data, contrarily to other tools, which only use radio data.\\
\emph{DeepSource} \citep{VafaeiSadr2019} is another CNN based solution that only detects point-sources, it does not classify objects, nor does it output a segmentation mask. 
The reported precision (recall) values on simulated datasets range from 0.45 (0.85) for an S/N of 3$\sigma$, to 0.99 (0.99) for S/N above 4$\sigma$.\\
\citealt{Mostert2022} employed Fast R-CNN architectures to perform radio-component association from the LOFAR Twometre Sky Survey (LoTSS) data, obtaining a level of accuracy ($\sim$84.0\%) comparable to that typically reached in crowdsourcing analysis.\\
In this context, it is important to highlight that all of these solutions are not directly comparable in performance as they were trained and tested on different data sets (real or simulated, different S/N levels, and data set sizes, etc.), targeting different types of objects, and producing different types of outputs.\\In this work we present a new source finding tool, named \toolname{} (Compact And Extended Source Automated Recognition with Mask R-CNN), aiming to tackle the discussed aspects in source extraction, using Mask R-CNN instance segmentation framework.
\emph{caesar-mrcnn} was trained to both detect and classify radio sources of different morphologies (compact or extended), imaging artefacts, and poorly imaged sources. The paper is organised as follows. In Section~\ref{sec:dataset-training} we describe the radio observations used, and the dataset produced for training and testing scopes.
In Section~\ref{sec:sfinder} we describe the Mask R-CNN object detection framework, and source finder implementation details. In Section~\ref{sec:results} we present the detection and classification results obtained on test radio images. Finally, future perspectives are reported in Section~\ref{sec:summary}.

\section{Dataset}
\label{sec:dataset-training}

\subsection{Observational data}
\label{sec:radio-surveys}

\subsubsection{ASKAP pilot surveys}
\label{subsec:askap}
The ASKAP EMU early science program was started in 2017, while the array commissioning was almost completed, to validate the array operations, the observation strategy, and optimize the data reduction pipeline. In this phase, several pilot surveys were carried out on target fields, also including the Galactic plane, bringing first scientific results. Among them, the EMU pilot survey \citep{Norris2021} (area=270 deg$^{2}$, rms=25–30 $\mu$Jy~beam$^{-1}$, angular resolution=$\sim$12.5"$\times$10.9" arcsec, central frequency=944 MHz) produced a source catalogue of $\sim$220,000 sources ($\sim$80\% single-component).\\
The \scorpio{} field (area=40 deg$^{2}$, centred on $l$=343.5$^{\circ}$, $b$=0.75$^{\circ}$) was the only field observed in the Galactic plane during the Early Science program, at different epochs, and in different ASKAP frequency bands. First observations at 912 MHz \citep{Umana2021} were carried out with 15 antennas, reaching an angular resolution of $\sim$24"$\times$21", and 200 $\mu$Jy~beam$^{-1}$ rms (far from the Galactic plane and bright sources). A compact source catalogue was reported in \cite{Riggi2021} for this field.\\Recent observations \citep{Ingallinera2022} with 36 antennas were done in all three ASKAP bands. The final total intensity map has a central frequency of 1243 MHz, $\sim$9.4"$\times$7.7" resolution, and $\sim$50 $\mu$Jy~beam$^{-1}$ rms. Due to the increased sensitivity, the number of detected compact sources increased by a factor 3.

\subsubsection{ATCA \textsc{Scorpio} survey}
ASKAP observations (see Section~\ref{subsec:askap}) complemented and significantly increased the field of view reached in previous observations of the \scorpio{} field ($\sim$8.4 square degrees), carried out with the Australia Telescope Compact Array (ATCA) at 2.1 GHz (rms$\sim$30-40 $\mu$Jy/beam, 9.8"$\times$5.8" angular resolution) \citep{Umana2015}, for which a compact source catalogue was reported in \cite{Riggi2021}.

\begin{table*}
\centering%
\scriptsize%
\caption{Number of images (Column 2) and object instance counts (Columns 3-10) per each class and radio telescope source (Column 1, see Section~\ref{sec:radio-surveys} for details) in the produced dataset. For the class "extended-multisland" we report the total number of objects (Column 5) and the number of objects with 2, 3 and more than 3 islands, respectively in Columns 6-8.}
\begin{tabular}{lccccccccc}
\hline%
\hline%
\multirow{3}{*}{\footnotesize{Telescope}} & \multirow{3}{*}{\footnotesize{\#Images}} & \multicolumn{8}{c}{\footnotesize{\#Objects}} \\%
\cmidrule(lr){3-10}%
& & COMPACT & EXTENDED & \multicolumn{4}{c}{EXTENDED-MULTISLAND} & SPURIOUS & FLAGGED\\%
\cmidrule(lr){5-8}%
& & & & All & 2 & 3 & >3 & &\\%
\hline%
VLA & 5780 & 5969 & 1740 & 1504 & 1180 & 315 & 9 & 4 & $-$\\%
ASKAP & 4090 & 15475 & 685 & 59 & 45 & 11 & 3 & 2276 & 286 \\%
ATCA & 2904 & 9089 & 924 & 120 & 85 & 33 & 2 & 206 & 5 \\%
\hline%
All & 12774 & 30533 & 3349 & 1683 & 1310 & 359 & 14 & 2486 & 291 \\%
\hline%
\hline%
\end{tabular}
\label{tab:dataset}
\end{table*}

\subsubsection{The Radio Galaxy Zoo (RGZ) dataset}
The Radio Galaxy Zoo (RGZ) \citep{Banfield2015} is a citizen science project where volunteers can classify radio galaxies and their host galaxies from radio and infrared images, made available through a web interface. More than 12,000 citizens contributed to the first Data Release (DR1) (Wong et al., in prep.), containing $\sim$75,000 sources. The radio data used in the project are mainly ($\sim$99\% of the sources) taken from the Faint Images of the Radio Sky at Twenty cm (FIRST) survey (1.4 GHz, angular resolution $\sim$5") \citep{Becker1995}, with only $\sim$1\% of the sources from the Australia Telescope Large Area Survey (ATLAS) \citep{Norris2006}.\\In this work we have used a subset\footnote{FITS images and relative annotation files available at \url{https://cloudstor.aarnet.edu.au/plus/s/agKNekOJK87hOh0}} 
of the RGZ DR1, produced by \cite{Wu2019}, and including only RGZ sources from VLA data (e.g. no sources from ATLAS survey were included) with consensus level $\ge$0.6 and with less than three components or peaks. Radio galaxies in the dataset are labelled into multiple classes, according to the observed number of components (C) and peaks (P): 1C-1P + 1C-2P + 1C-3P (68\%), 2C-2P + 2C-3P (21\%), 3C-3P (11\%).

\subsection{Source labeling scheme}
\label{subsec:class-label-schema}
\begin{figure*}
\centering%
\subtable[ASKAP]{\includegraphics[scale=0.18]{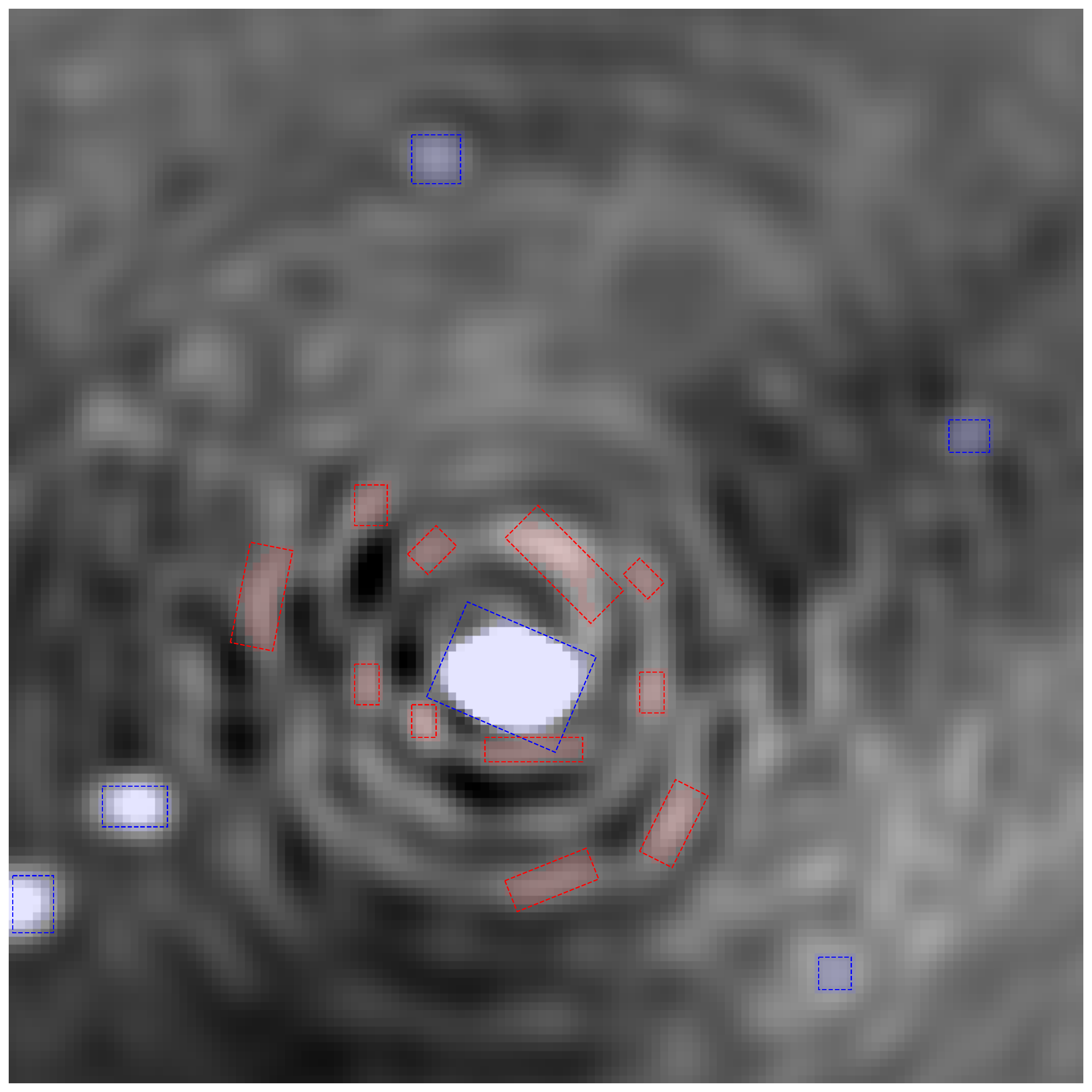}}%
\subtable[ATCA]{\includegraphics[scale=0.18]{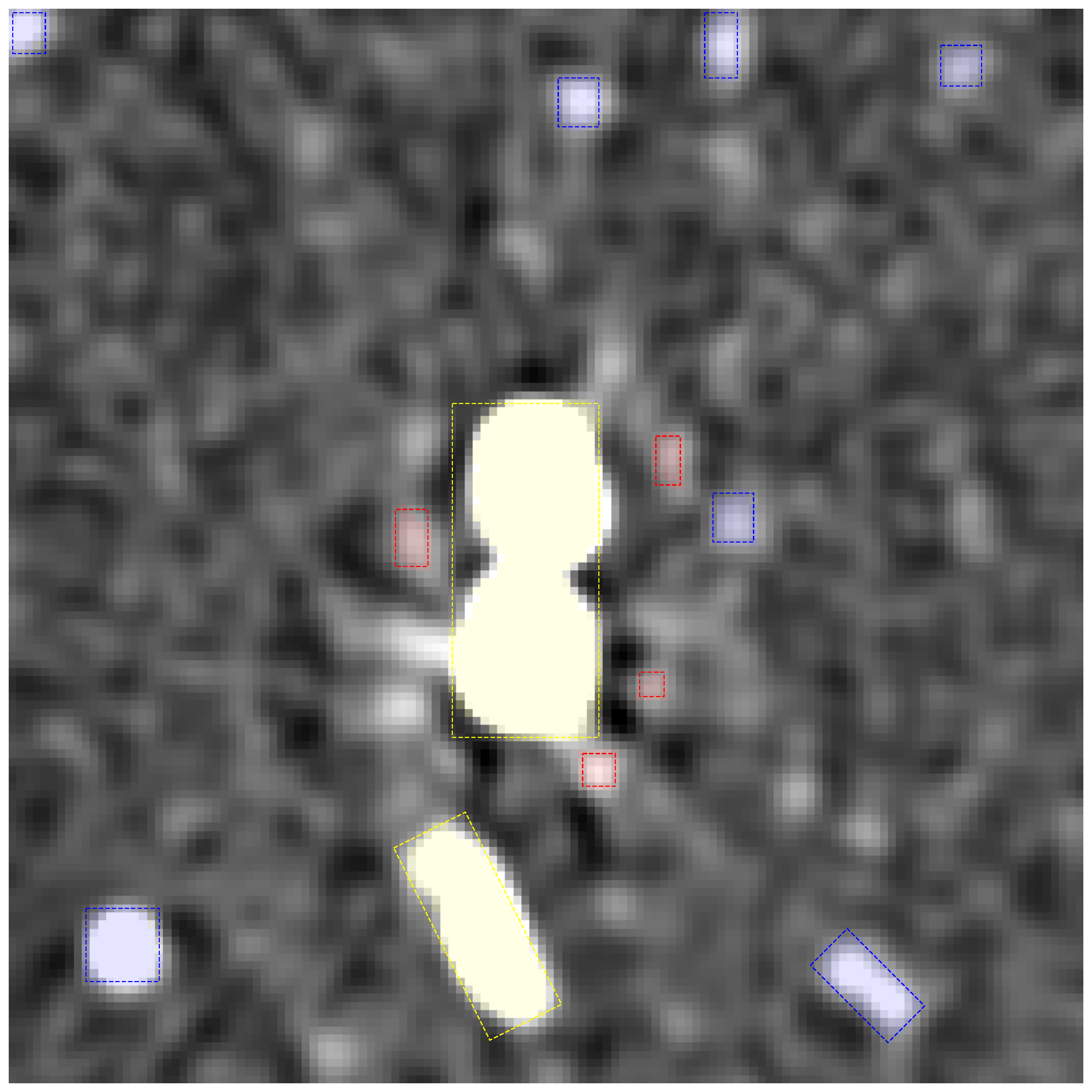}}%
\subtable[ASKAP]{\includegraphics[scale=0.18]{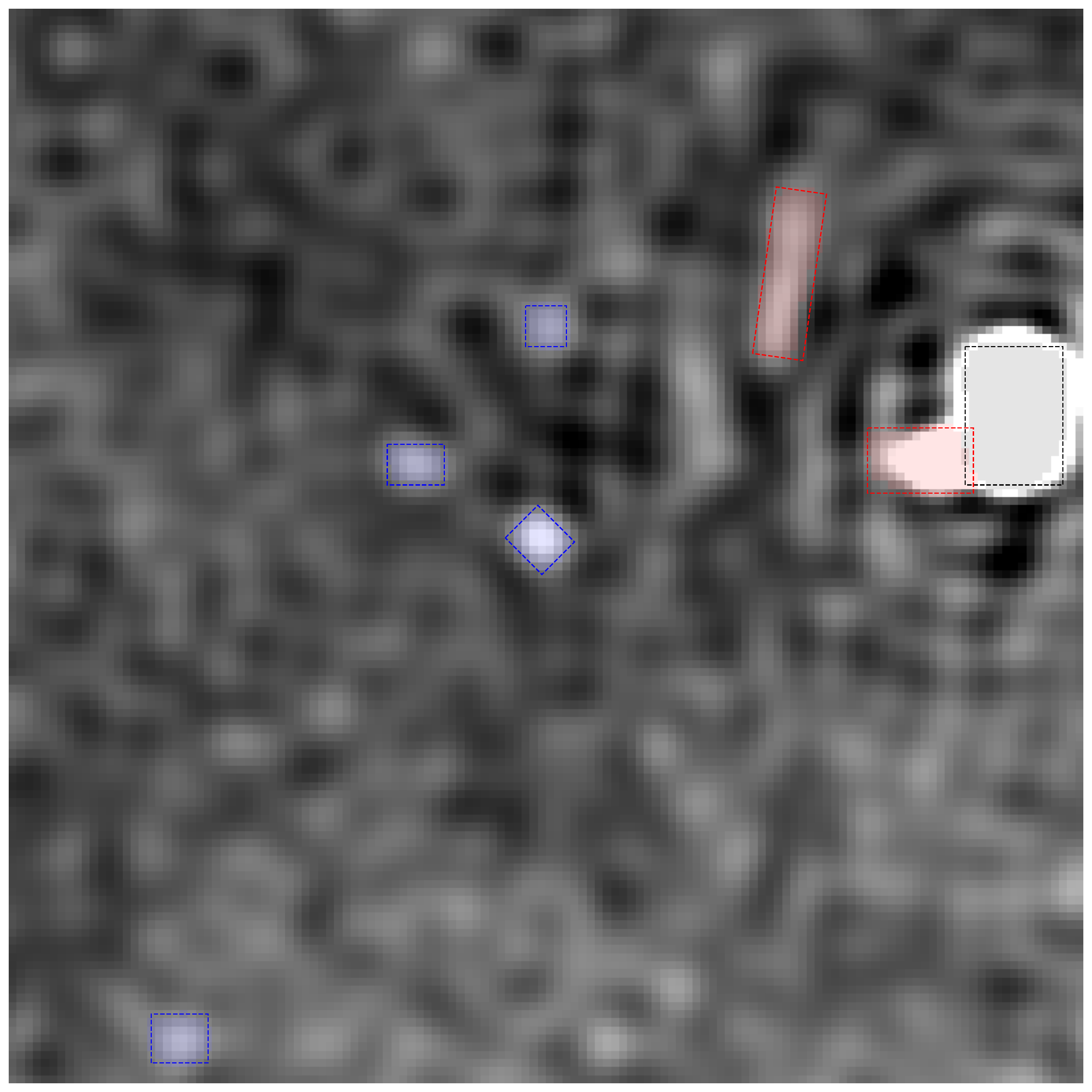}}\\%
\subtable[ATCA]{\includegraphics[scale=0.18]{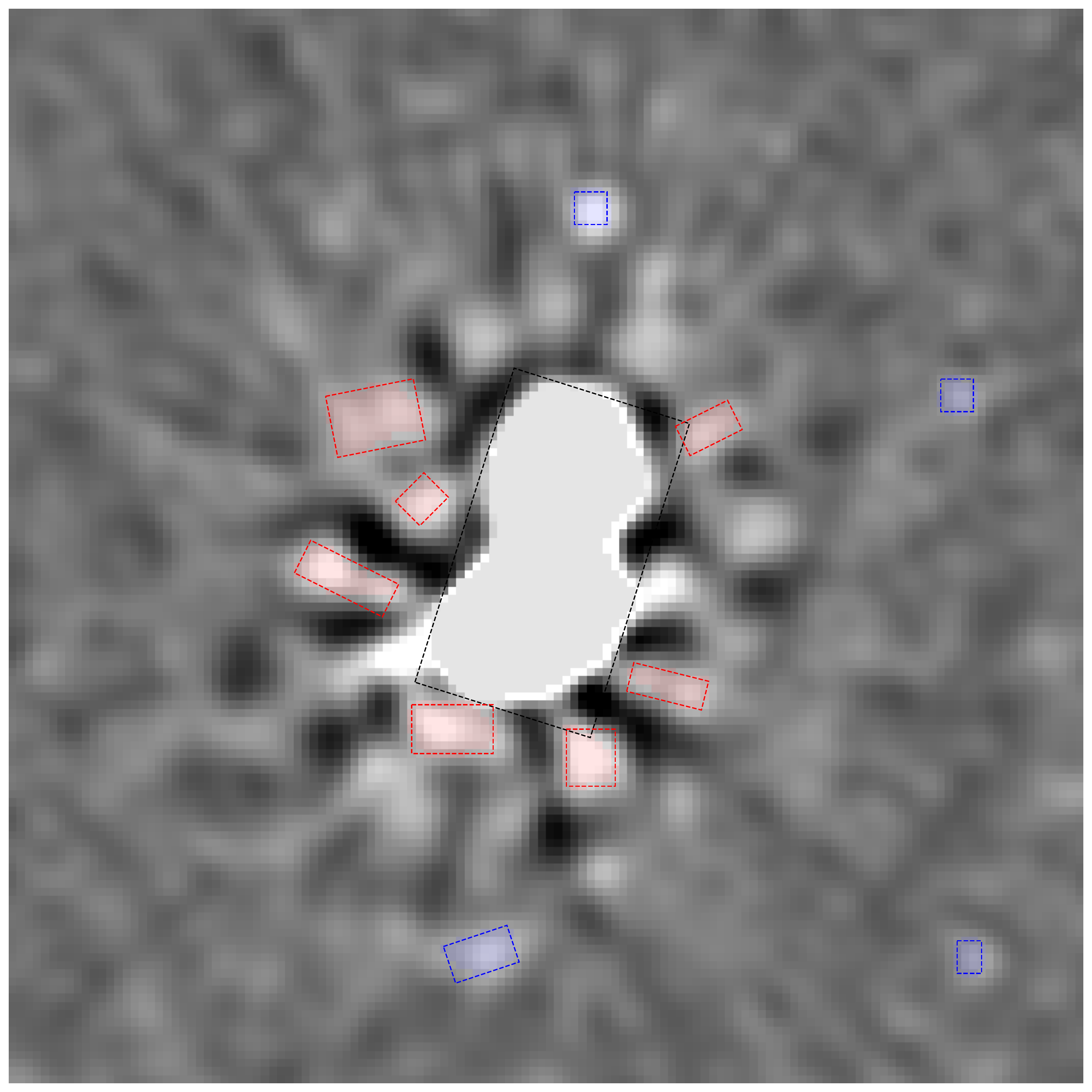}}%
\subtable[ASKAP]{\includegraphics[scale=0.18]{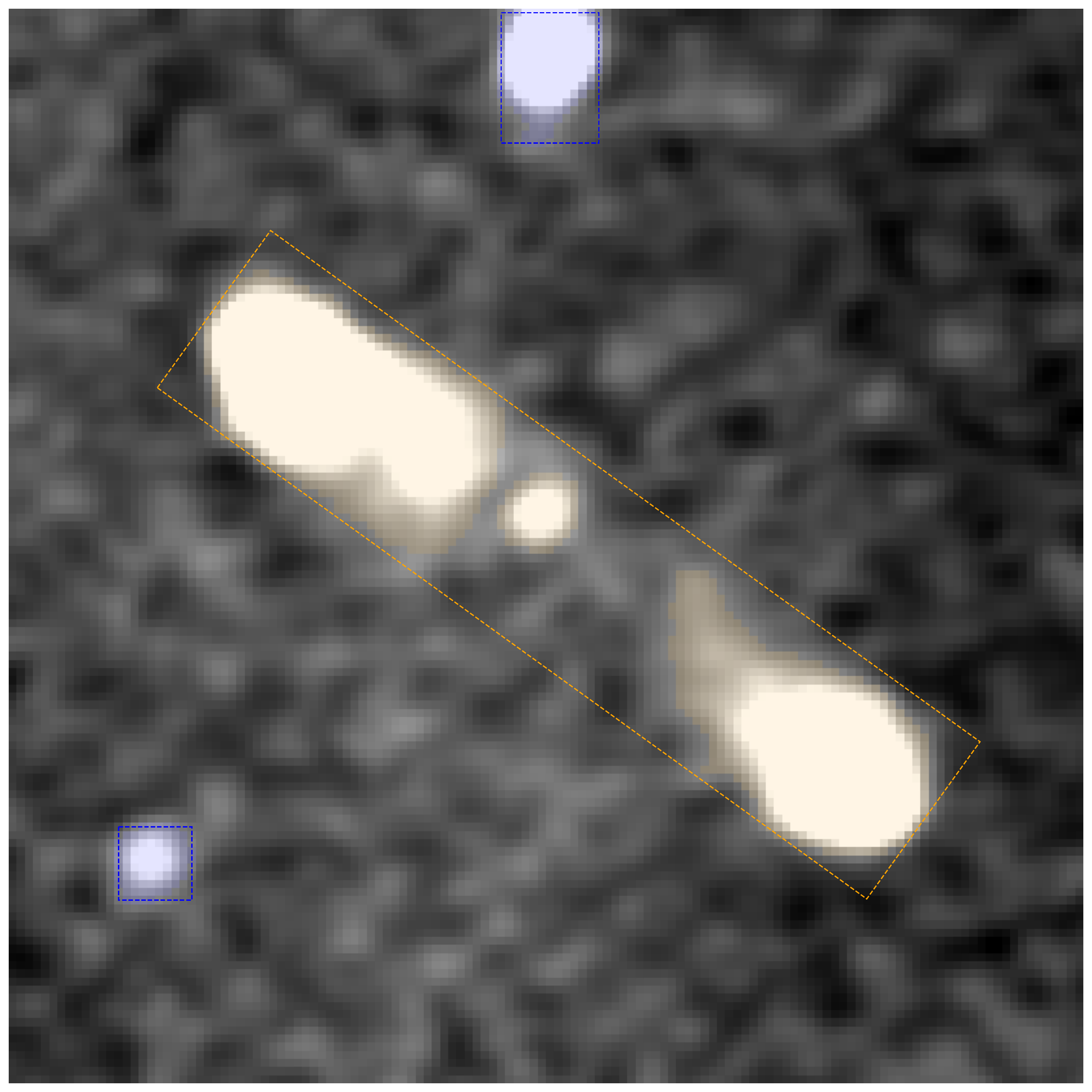}}%
\subtable[ASKAP]{\includegraphics[scale=0.18]{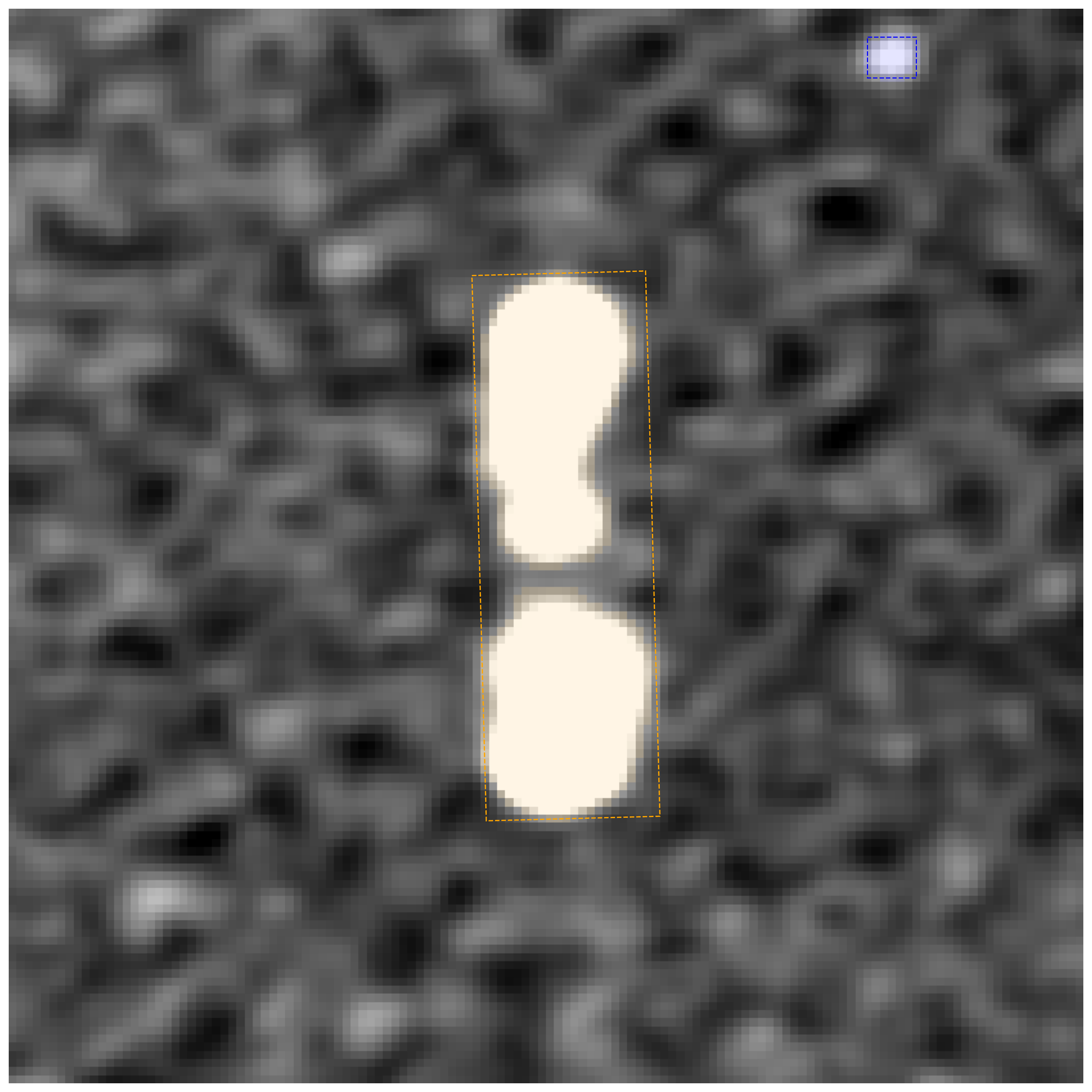}}\\%
\subtable[VLA]{\includegraphics[scale=0.18]{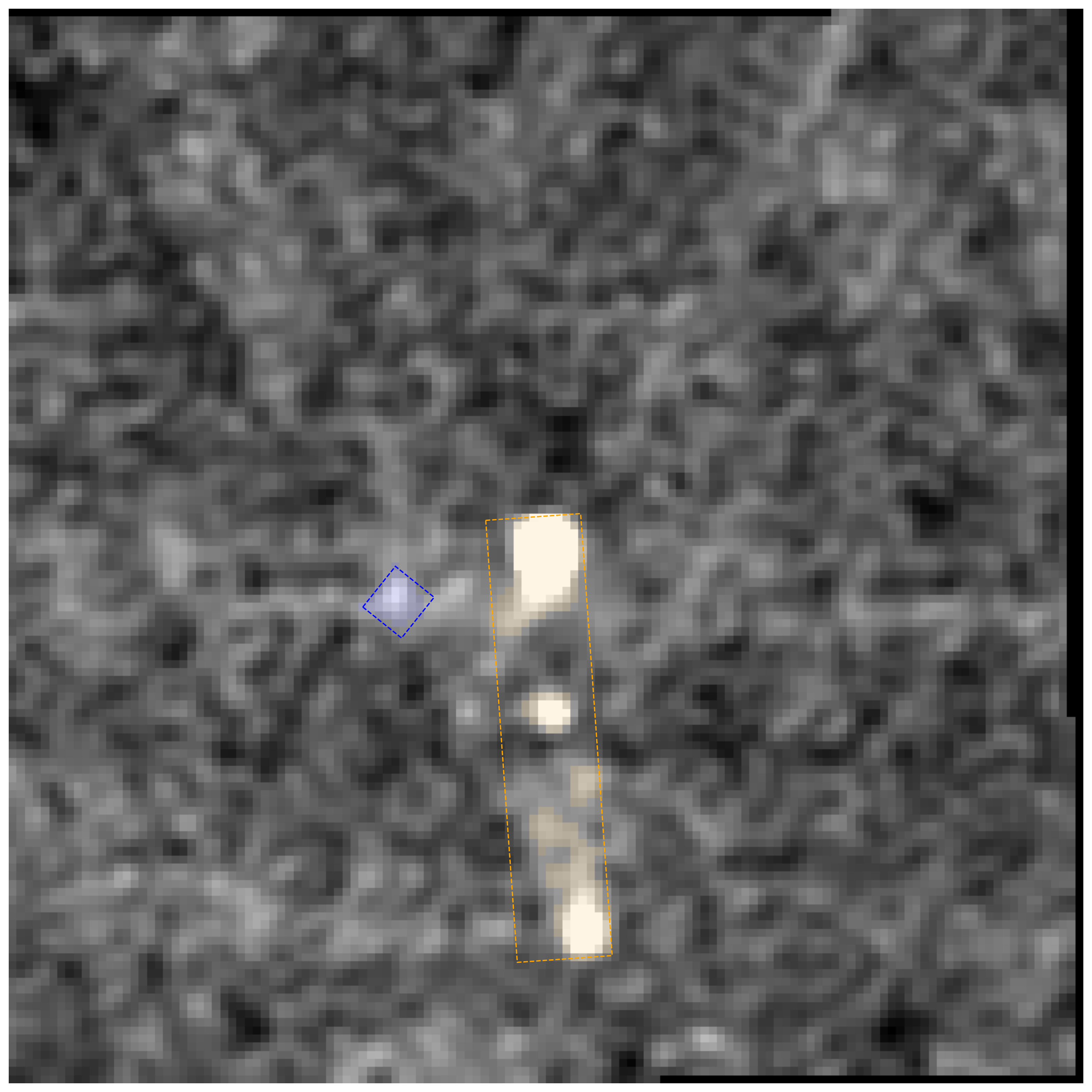}}%
\subtable[VLA]{\includegraphics[scale=0.18]{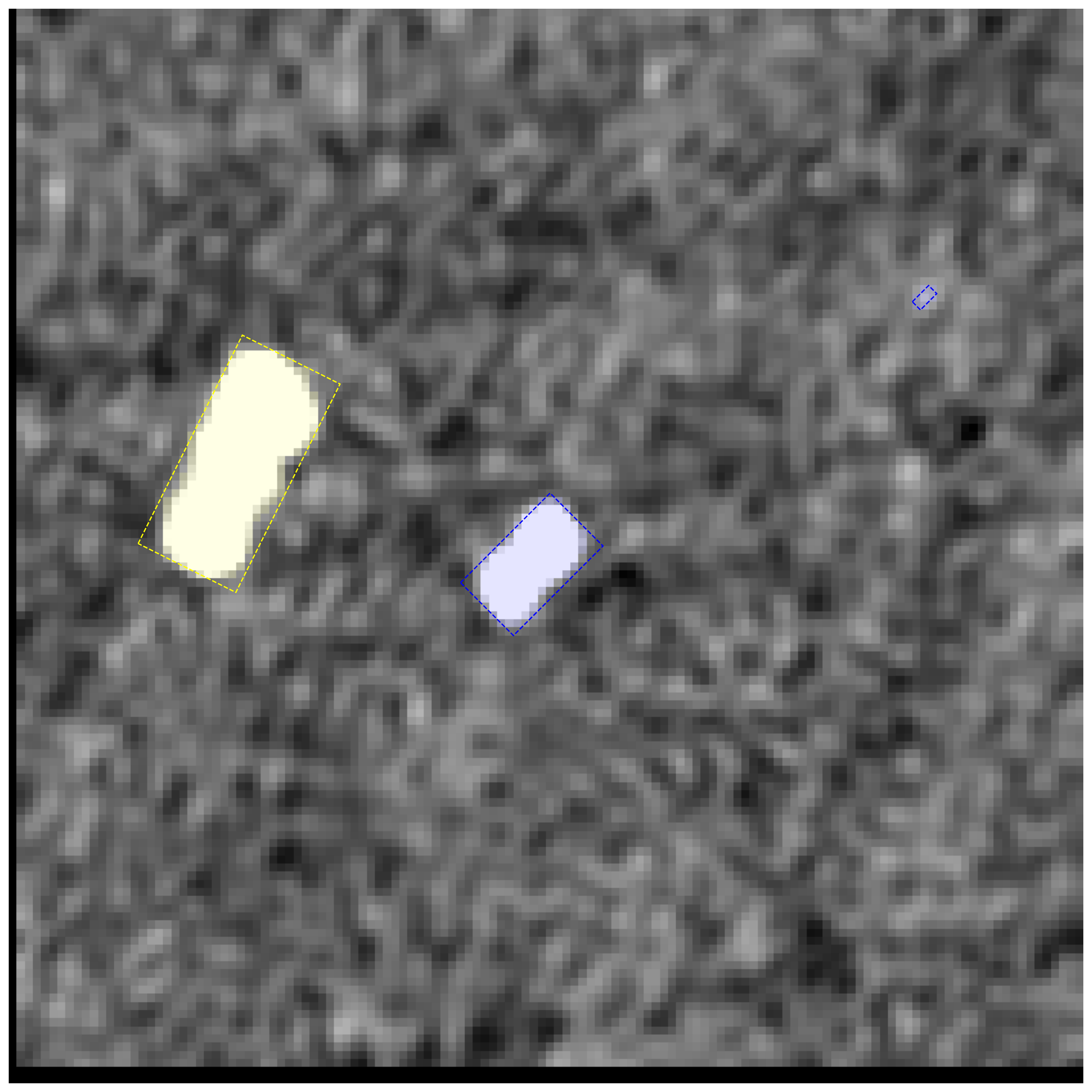}}%
\subtable[ASKAP]{\includegraphics[scale=0.18]{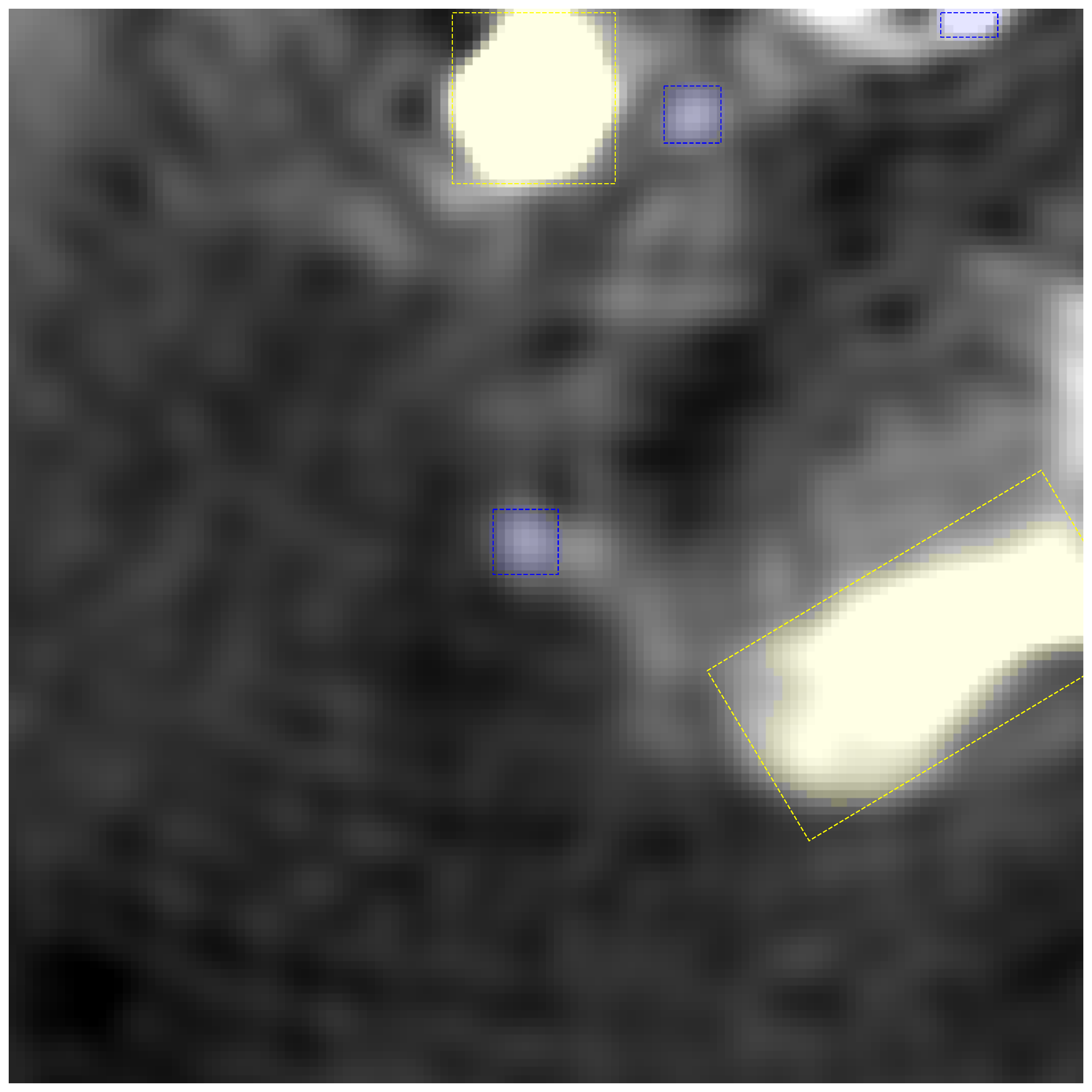}}%
\caption{%
Sample radio images from the dataset with ZScale transform applied (see Section~\ref{sec:preprocessing}), and rotated bounding boxes of all labelled objects superimposed: compact sources are shown in blue, extended in green, extended-multisland in orange, spurious sources in red, and flagged sources in gray. Below each panel, we report the image telescope source.
}%
\label{fig:sample-inputs}
\end{figure*}

In accordance with the scientific use case described in Section~\ref{sec:intro}, we considered five object classes as the targets for our source finder:
\begin{enumerate}
\item \texttt{\textsc{compact}}: including single-island isolated point- or slightly resolved compact radio sources, eventually hosting one or more blended components, each with morphology resembling the synthesized beam shape;
\item \texttt{\textsc{extended}}: including radio sources with a single-island extended morphology, eventually hosting one or more blended components, with some deviating from the synthesized beam shape;
\item \texttt{\textsc{extended-multisland}}: including radio sources with an extended morphology, consisting of more (point-like or extended) islands, each one eventually hosting one or more blended components; 
\item \texttt{\textsc{spurious}}: including spurious sources, due to artefacts introduced in the radio map by the imaging process. Artefacts can be of different types, often with patterns following the UV coverage of the array. We are limiting here to sidelobes around bright sources, which have a ring-like or elongated compact morphology. They are typically also bright, thus passing the 5$\sigma$ significance threshold applied by many traditional source finders\footnote{Some source finders, like PyBDSF, have recently implemented strategies to minimise sidelobe extraction, based on an improved rms noise estimation close to bright sources.}, and reducing the catalogue reliability even at high S/N levels;
\item \texttt{\textsc{flagged}}: including single-island radio sources, with compact or extended morphology, that are poorly imaged and cannot be separated from close imaging artefacts. These sources are typically very bright, and are to be excluded from the source catalogue, or at least flagged, as their properties (e.g. flux density, shape) cannot be reliably measured.
\end{enumerate}
For analysis scopes, we also define a parent class \texttt{\textsc{SOURCE}}, including real and non-flagged sources, i.e. object instances of class \texttt{\textsc{compact}}, \texttt{\textsc{extended}}, or \texttt{\textsc{extended-multisland}}.
Sample images from the dataset, with labelled objects superimposed, are reported in Figure~\ref{fig:sample-inputs}.\\
We are deliberately adopting a simple and conservative labelling scheme, that is science-agnostic (e.g. the astrophysical nature of each source is not considered nor forced), only relying on the radio continuum (e.g. no comparison with other wavelengths involved), and using broad morphological categories that can be widely understood and adopted in different radio domains as a first-order tagging scheme before more refined classification stages are applied (e.g. typically employing unsupervised or self-supervised methods). Our first goal is, in fact, to produce a model that can identify radio sources (irrespective of their morphology, e.g. compact, single- or multi-island extended), from artefacts and poorly imaged sources. Then, as a secondary step, we would like the model to provide indications about the source morphology, that can be eventually used as inputs for new classification algorithms.\\Furthermore, for this work, we did not consider diffuse sources, e.g. extended sources without sharp edges typically found along the Galactic plane. We, however, plan to include them as a new object class in a future version of the dataset.

\subsection{Dataset preparation}
\label{subsec:dataset-preparation}
To build our dataset, we searched for sources of all classes, as defined in the previous paragraph, in the available radio data described in Section~\ref{sec:radio-surveys}.
As the data preparation process to label, segment and inspect sources require significant efforts, we did not use the full image and source catalogue samples available for some observations.\\Input image cutouts (single-channel, size 132$\times$132 pixels\footnote{The source cutout size in pixels was chosen equal to that used in the RGZ dataset, corresponding to a 3 arcmin$\times$3 arcmin field of view for a pixel size of 1.375" in FIRST images. The ratio between the image psf and pixel size is rather comparable among the different surveys used in the dataset (VLA=3.6, ASKAP=5, ATCA=3.9) so we do not expect significant discrepancies for compact source detection due to different psf sampling.}, FITS format) were extracted from the reference data. To train our source detection framework, a series of mask images are required for each input image, each of them containing the object binary segmentation. A raw object segmentation was preliminarily produced with the \caesar{} source finder \citep{Riggi2016,Riggi2019} for each cutout and later refined by visual inspection using a collaborative approach.
\begin{figure*}
\centering%
\subtable{\includegraphics[scale=0.31]{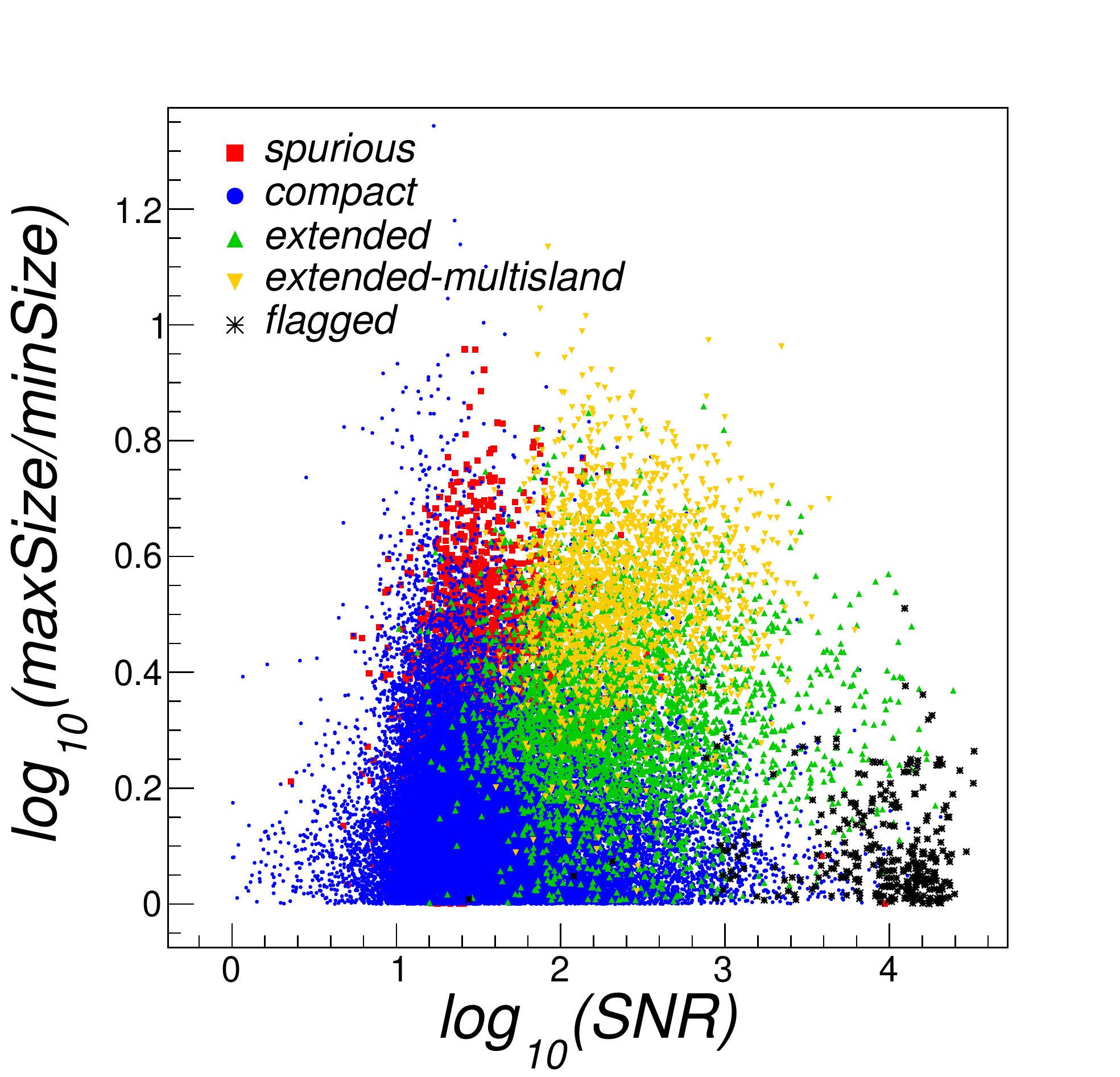}}%
\hspace{-0.4cm}%
\subtable{\includegraphics[scale=0.31]{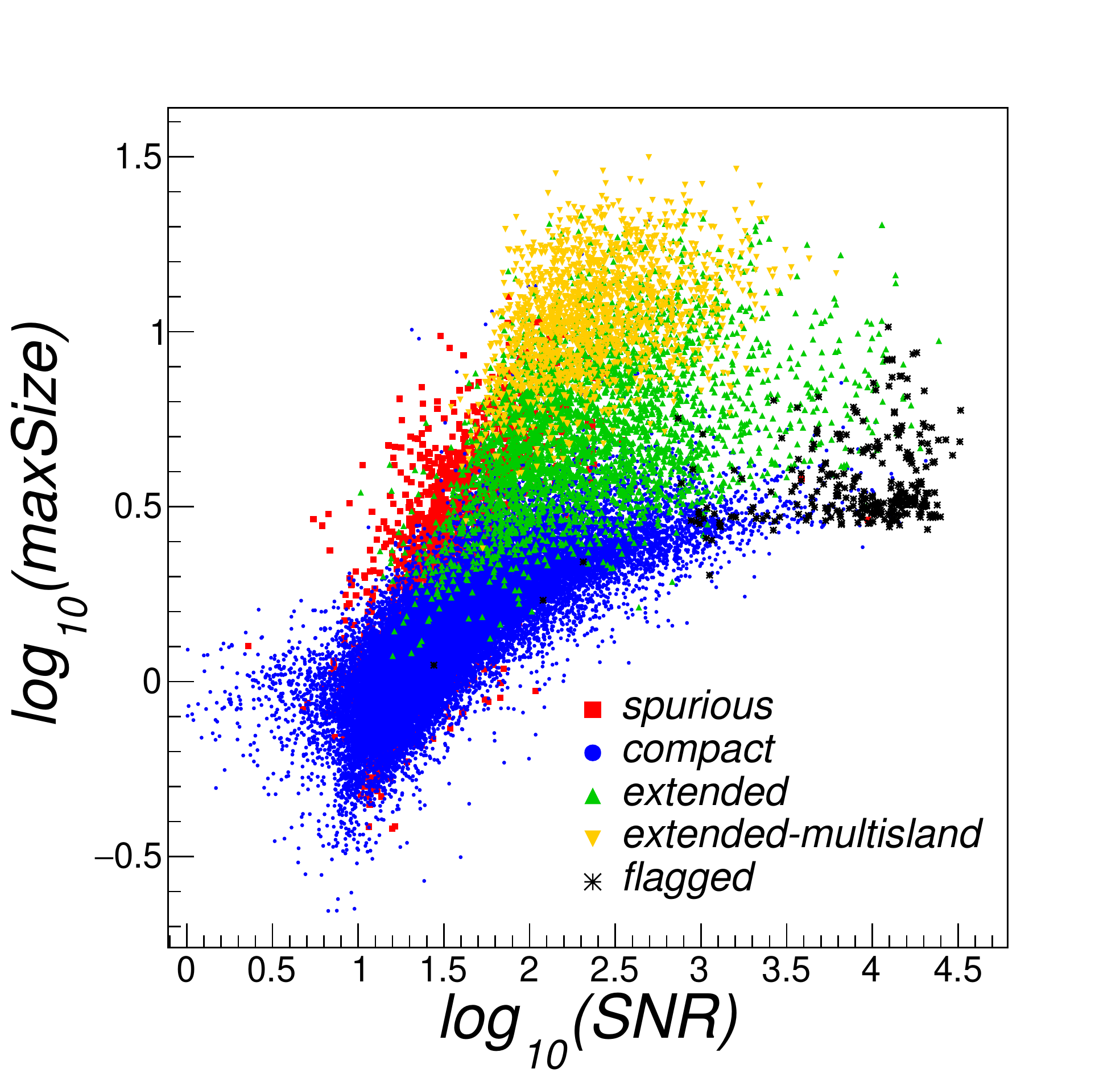}}%
\hspace{-0.4cm}%
\subtable{\includegraphics[scale=0.31]{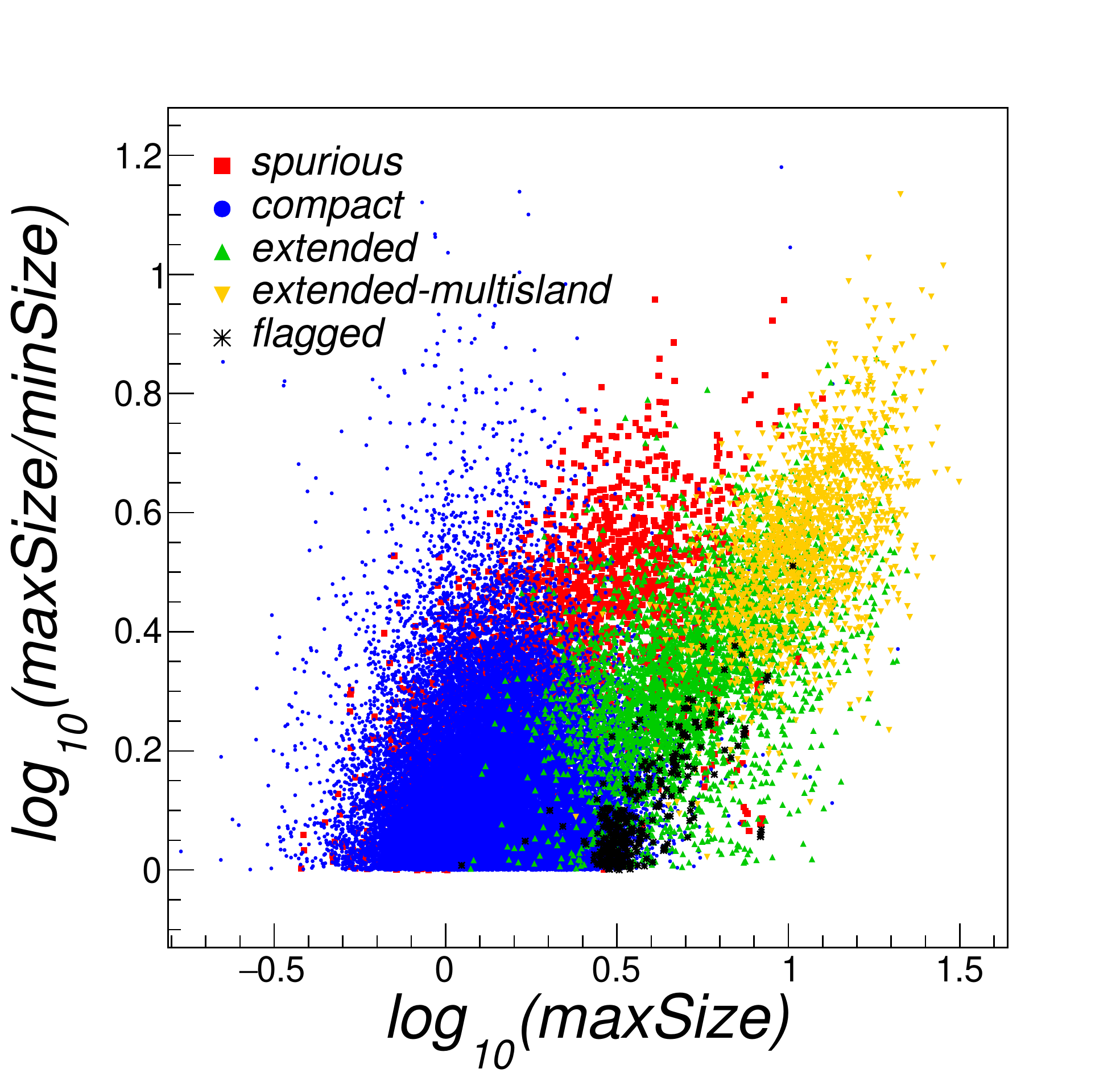}}%
\caption{Scatter plots of the source signal-to-noise ratio (SNR), maximum source size and aspect ratio for all object classes. See text for parameter definition.}%
\label{fig:scatter-plots}
\end{figure*}
\caesar{} default parameter configuration (see \citealt{Riggi2019}) worked in the majority of cases for point sources, without requiring manual refinement of the mask. This is not always the case for extended sources, in which the diffuse lobe emission is often missed by the finder at the standard 5$\sigma$ threshold. A refinement is also typically needed to separate bright sources and their surrounding artefacts, that are often blended and extracted by the finder as a unique island. Finally, fainter sources missed in the automated process were manually added to the list of labelled objects when clearly distinguished from the background noise.\\
Raw data (input images, segmentation regions in DS9 format) were first ``anonymised'' (e.g. WCS parameters and any observatory reference metadata were removed from the image header) and then made available to a data preparation team in a git repository. The team was divided into different groups (3 or 4 people per group), each of them manually inspecting and revising the classification labels and pixel segmentation for a subset of the raw data. Masks are, finally, automatically produced from the improved segmentation regions. Both raw and final data are kept under version control during the process, using the Data Version Control (\textsc{dvc}) framework\footnote{\url{https://dvc.org/}}.\\
Results and source counts obtained per each class and reference telescope are summarized in Table~\ref{tab:dataset}. The final, accumulated, dataset consists of $\sim$12,700 images, comprising $\sim$30,500 compact sources, $\sim$3,300 extended sources, $\sim$1,600 multi-island extended sources, $\sim$2,400 spurious sources, and $\sim$290 flagged sources. The number of available images is not severely unbalanced with respect to the telescope source: $\sim$45\% (VLA), $\sim$32\% (ASKAP), $\sim$23\% (ATCA). Compact sources are the most abundant objects in the sample ($\sim$80\%), followed by extended single- or multi-island sources ($\sim$13\%), which are mostly ($\sim$64\%) obtained from VLA observations\footnote{The predominance of VLA extended sources is at present mostly due to the limited area covered by some ASKAP/ATCA surveys (e.g. the \textsc{Scorpio} field survey) compared to FIRST survey coverage, and to the limited labelling efforts spent so far in those surveys compared to RGZ crowd-sourcing.}. The number of available spurious and flagged sources is smaller than 7\%, largely obtained from ASKAP observations. Future versions of the dataset should aim to reduce the class unbalances reported above.\\
In Fig.~\ref{fig:scatter-plots} we report the distribution of each object class as a function of the computed source signal-to-noise ratio (SNR), maximum source size (e.g. the largest dimension of the source rotated bounding box, expressed as multiple of the synthesized beam size), and aspect ratio (e.g. the ratio between the larger and the smaller rotated bounding box dimensions). Labelled compact sources have a roundish shape in most cases (single-component), and SNRs decreasing below the 5$\sigma$ threshold. Spurious sources have SNRs well above the 5$\sigma$ threshold, and an elongated shape in most of the cases. Extended sources are, as expected, the largest and most elongated objects in the dataset. Flagged sources are the brightest objects in the dataset. Many have a roundish shape, much larger than the synthesized beam shape.

\subsection{Data pre-processing}
\label{sec:preprocessing}
A pre-processing step was applied to the images before being given as classifier inputs. Any `NaN' pixel value was first set to the image minimum (e.g. the smallest finite value) and pixel values were normalised using a ZScale transform \citep{NOAO1997ZScale}. Finally, pixel values were normalised in the range [0,255] and input images transformed to 3-channel RGB (by replicating the single gray level channel) as the original model architecture was designed to work with RGB images. No background subtraction step was applied to the resulting images, as the network is expected to learn the noise pattern from the data.\\
During training only, image augmentation is used to prevent the model from overfitting. A random number of augmentation steps (0 to 2) are applied to each image before being processed by the model. The augmentation operations implemented are: flipping left to right and upside down, rotating $90\deg$ clockwise or anti-clockwise, and image translations. How many and which operations are applied to a particular image are random and different each epoch.

\begin{figure*}[h]
\centering
\includegraphics[width=0.85\textwidth,keepaspectratio]{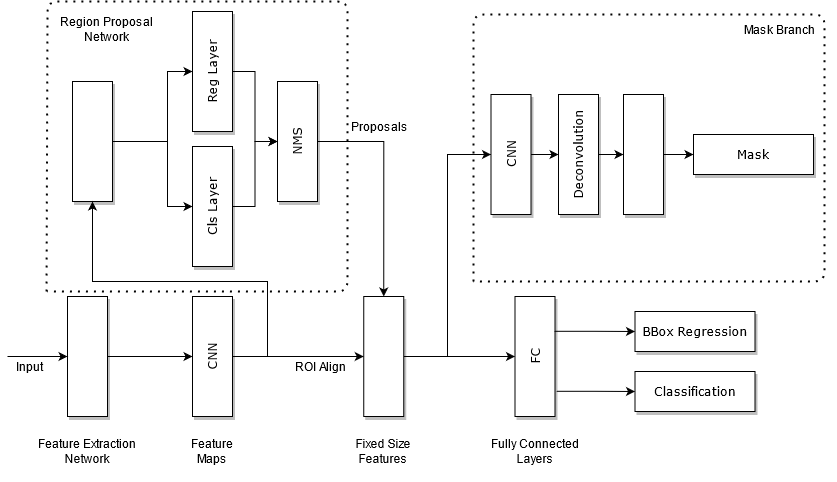}
\caption{Mask R-CNN architecture, reproduced from \cite{Yang2020}.}
\label{fig:MaskRCNNarch}
\end{figure*}

\section{\emph{caesar-mrcnn}}
\label{sec:sfinder}

\subsection{Mask R-CNN: A Framework for Object Detection and Classification}
\label{subsec:mrcnn}
Mask R-CNN is a deep learning model that has been recently proposed, combining the capability of performing object detection, classification, and instance segmentation on images. The algorithm was developed by the Facebook AI Research team in 2017 and has been used in many computer vision problems such as identifying vehicles or faces~\citep{He2017}, marine mammals~\citep{gray2019}, and astronomical optical sources~\citep{burke2019deblending}. Mask R-CNN is built upon, and shares a similar architecture to its predecessor Faster R-CNN~\citep{Ren2017}. The additions include a Feature Pyramid Network (FPN) as part of the backbone, replacing the region of interest (ROI) pooling step with a ROI align, and the addition of a fully convolutional network (FCN) for predicting mask instances. Fig.~\ref{fig:MaskRCNNarch}, reproduced from \cite{Yang2020}, shows a high-level schematic of the Mask R-CNN architecture.

\subsubsection{Feature Pyramid Network}
In the first Mask R-CNN stage, an image is passed to a Feature Pyramid Network, which serves as the backbone architecture for extracting object features at multiple scales. The FPN is built on two pyramid-like pipelines called the bottom-up pathway and the top-down pathway \citep{Lin2017}.

The bottom-up pathway uses a Residual Network (ResNet) having 5 stages (a pyramid level) with Residual Blocks \cite{He2016} to generate a number of feature maps each with varying filters. The number of Residual Blocks and their depth depends on the size of the network.

The top-down pathway traverses the second pyramid by picking the highest feature map from the bottom-up pathway as its highest map. These undergo a $1\times1$ convolution to reduce the channel dimensions. Lateral connections, similar to skip connections in ResNet, add the upsampled feature map to the $1\times1$ convolution to create the next feature map (\citealt{Lin2017}, Fig. 3). This process continues for all layers in the corresponding bottom-up pathway. To create the final feature maps, each layer in the top-down pathway undergoes a $3\times3$ convolution. Additionally, a fifth feature map is created from a MaxPool operation on the highest feature map from the top-down pathway and is only used during the Region Proposal Network (RPN). This process allows these semantically strong features to be present at any level of the hierarchical structure, and lower resolution layers to detect features like edges and higher levels detecting objects~\citep{He2016}. The feature maps are then passed to the Region Proposal Network (RPN) to determine the Regions of Interest (ROI).

\subsubsection{Region Proposal Network}
The RPN is a network that scans over regions on the feature maps called anchors, a concept introduced in Faster R-CNN~\citep{Ren2017}, which replaces the selective search algorithm in Fast R-CNN~\citep{Girshick2015}. Many anchors (configurable in number) are suggested with predetermined sizes with respect to the original image, and aspect ratios (to cover for various box/rectangle ratios) are scanned in parallel. The RPN has a classifier to predict whether the region is either foreground or background (with an objectness score). Aside from classification, objects are localised through bounding box regression. Each object therefore is described by bounding box offsets and a class label.

Results from this classifier are sorted by objectness score, and only a specific number of anchors with the highest objectness score are kept. Furthermore, anchors with incorrect sizes or coordinates (negative anchors) are dropped. Another pruning step is the consideration of the Non Maximum Suppression (NMS) parameter which is a predefined ratio that compares the Intersection over Union (IoU) between overlapping anchors, removing ones with an IoU greater than the threshold. The net effect is that duplicate object detection anchors are removed. The remnant region proposals, called ROIs are passed on to the next stage.

\subsubsection{Bounding Box and Mask Branches}
Prior to bounding box branch processing, each ROI is mapped to their corresponding feature map, and will need ROIAlign to resize the dimensions to a defined pool size. The task of ROIAlign is to grid the ROI into equal segments, applying bilinear interpolation to each. The application of ROIAlign has shown to improve mask predictions compared to the previous technique of ROIPool in Faster R-CNN, which suffered from poor mask quality and regular spatial misalignment of mask pixels. 

With every ROI pooled to the same size, ROIs are passed to a fully connected layer for object classification and another fully connected layer for the final bounding box regression. These optimised bounding boxes are used in the mask branch to produce the mask predictions. Similar to the Bounding Box Branch, the ROIs also need to be passed through the ROIAlign process. At the end of this mask branch, ROIs are fed into a FCN which outputs a per-pixel prediction, coupled with a scaled 28x28 mask for each instance. The FCN, first introduced in 2015, contains 4 stages of $3\times3$ convolutions, Batch Normalization and ReLu activation layer to classify the pixels in the ROIs~\citep{long2015}. A deconvolution layer is used to resize the image to $28\times28$. A final $1\times1$ convolution with a sigmoid activation is used to classify object instances into their corresponding classes.

\subsubsection{Loss Calculation}
\label{subsec:loss-calculation}
The multitask loss equation calculated during training for each ROI is computed as:
\begin{equation}
L = L_{class} + L_{mask} + L_{box}
\label{eqn:loss_calc}
\end{equation}

$L_{class}$ is calculated as follows for the RPN, describing the ability of the classifier to separate between foreground and background, and for the ROI classifier in stage two:
\begin{equation}
L_{class} = - \log p_u 
\label{eqn:L_class}
\end{equation}
where $p_u$ is the probability of an anchor/ROI belonging to true class label $u$.

Similarly to $L_{class}$, $L_{box}$ also has two components and describes the box branch's ability to localise objects during the bounding box regression:
\begin{equation}
    L_{box} = \sum_{i=1} L_{1}^{\text{smooth}}\left(t_{i}^{u} - v_{i}\right)
\label{eqn:l_box}
\end{equation}

\begin{equation}
L_{1}^{\text{smooth}}(x) = \left\{\begin{array}{lr}
        0.5x^2, & \text{if } |x| < 1\\
        |x| - 0.5, & \text{otherwise}
        \end{array}\right\}
\label{eqn:smooth_l1}
\end{equation}
where $t_{i}^{u}$ and $v_{i}$ are the predicted and ground truth bounding box coordinates in pixels, respectively.

$L_{mask}$ is described as the average binary cross entropy per pixel and describes the mask heads' ability to classify each pixel value in a mask of size $m{\times}m$ for each class K:
\begin{equation}
L_{mask} = - \frac{1}{m^2} \sum_{i,j = 1}^{m} [y_{ij}^{K} \log \hat{y}_{ij} + \left( 1-y_{ij}^{K}\right)\log \left(1-\hat{y}_{ij} \right)]
\label{eqn:l_mask}
\end{equation}

The final outputs of Mask R-CNN are optimal bounding box detections for each object in an image, the corresponding class label together with a probabilistic confidence, as well as a binary mask for each object instance.

\begin{figure*}
\centering%
\includegraphics[scale=0.185]{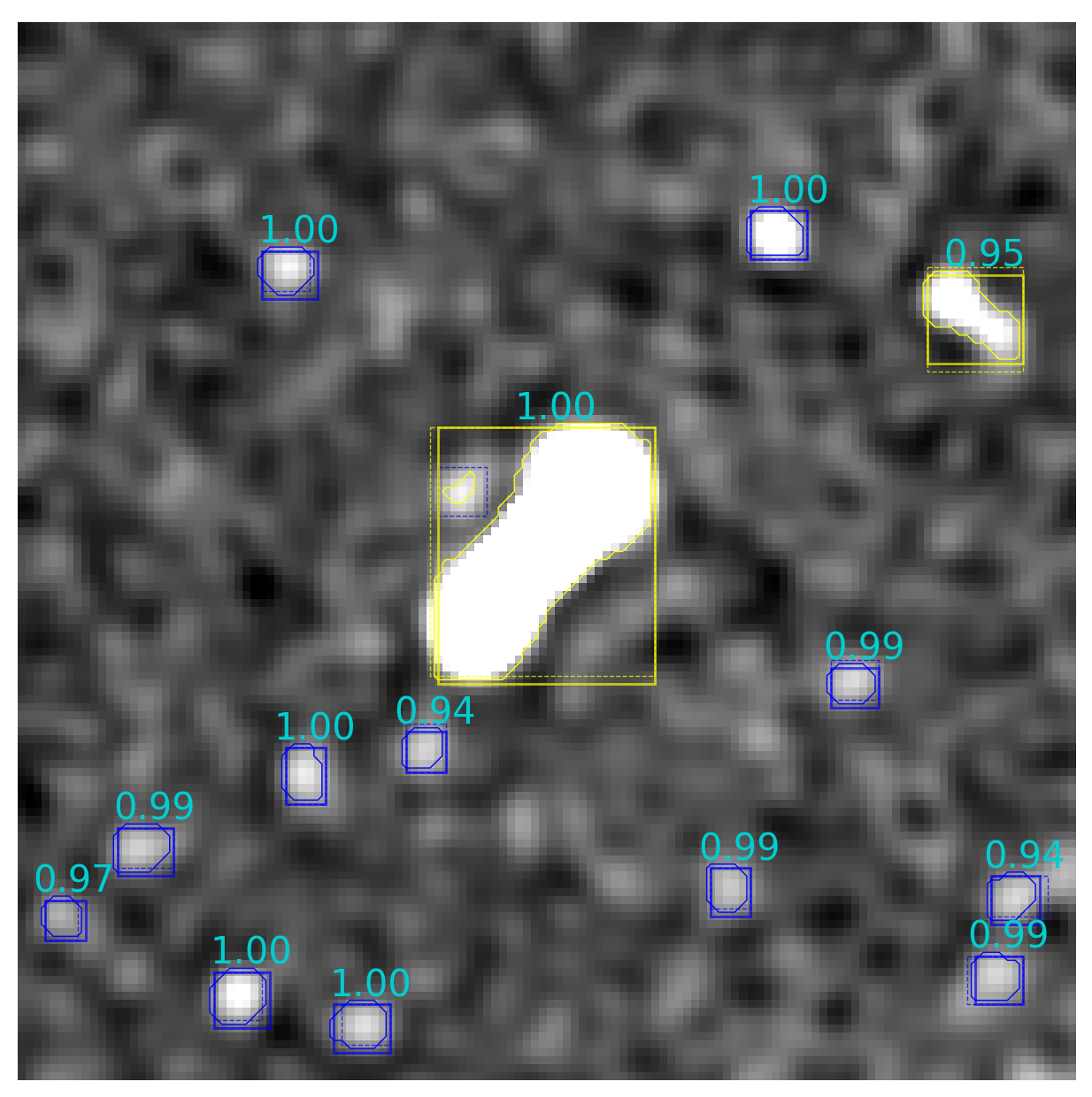}%
\hspace{0.3cm}%
\includegraphics[scale=0.185]{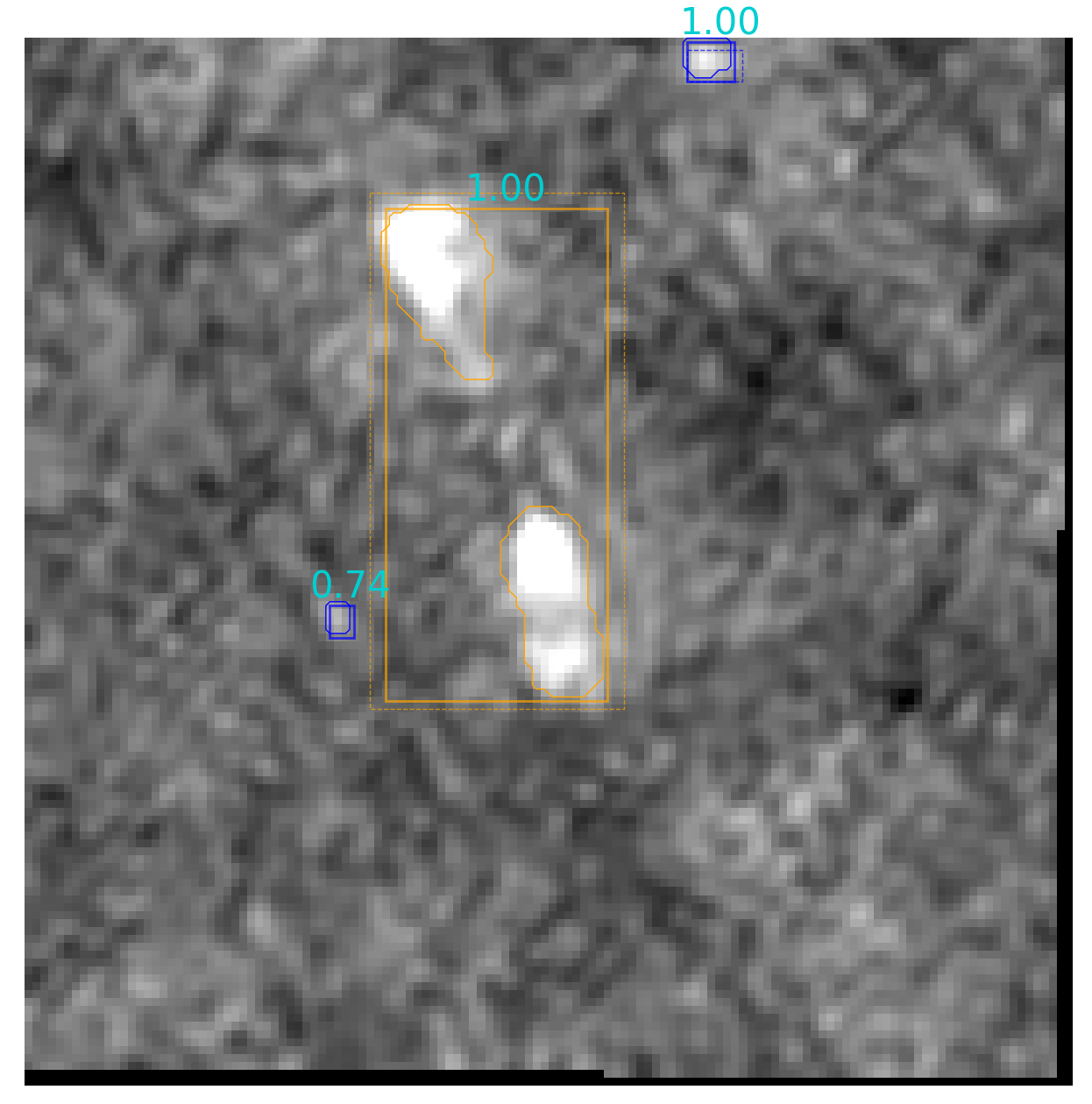}%
\hspace{0.3cm}%
\includegraphics[scale=0.185]{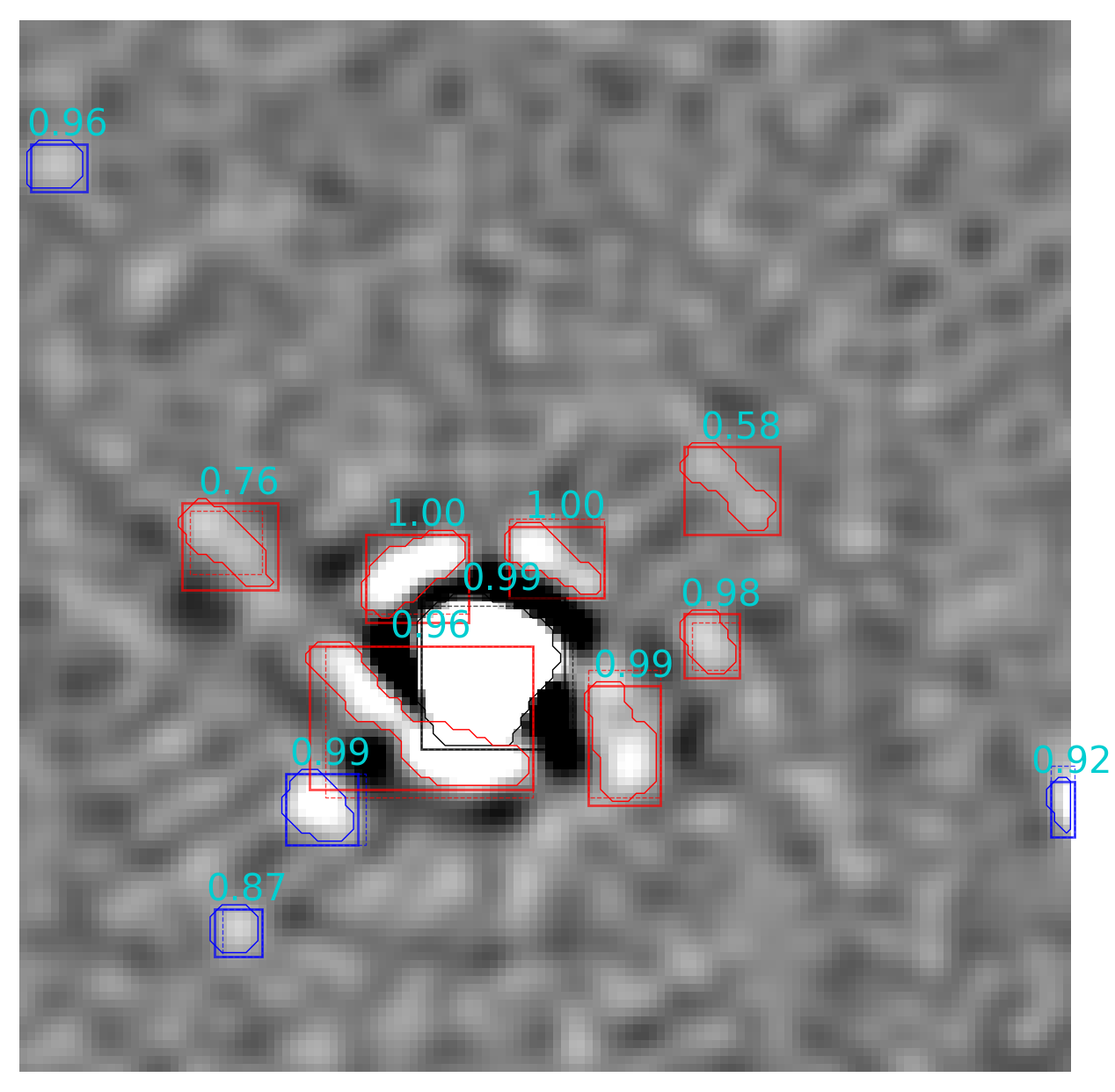}%
\caption{%
Sample \toolname{} outputs (bounding boxes, segmentation masks and classification score of detected objects) on three sample images from our dataset. Left panel: compact (blue) and single-island extended sources (yellow), taken from ATCA data; Centre panel: 2-island extended source (orange) and compact sources (blue), taken from RGZ data; Right panel: flagged source (gray) surrounded by imaging artefacts (red) and non-flagged compact sources (blue), taken from ASKAP data.
}%
\label{fig:sample-output}
\end{figure*}

\subsection{Model implementation}
\label{subsec:model}
\toolname{} is based on a Mask R-CNN implementation\footnote{\url{https://github.com/matterport/Mask_RCNN}} written in python and using the TensorFlow \citep{Abadi2015} and Keras \citep{Chollet2015} libraries.
We made, however, a number of modifications and extensions with respect to the original implementation, described in the following paragraphs and publicly available at \url{https://github.com/SKA-INAF/caesar-mrcnn}.

\subsubsection{Data loading}
\label{subsec:data-loader}
As the original Mask R-CNN implementation only supports RGB images (e.g. PNG format, usually) as inputs, we added a data loader for 2D FITS images, the typical format of radio continuum data, as well as a series of pre-processing options to transform the scale of input data before training. In this case, as pre-trained models and weights are only available for 3-channel images, the data loader reshapes the input image to 3 channels by replicating the single grayscale channel. Nevertheless, we also added support for training the model from scratch directly on single channel (grayscale) images.\\Input images can be either small cutouts, e.g. with size comparable to the Mask R-CNN \texttt{IMAGE\_MAX\_DIM} parameter (typically 256, 512 or 1024 pixels per side), or larger images if the parallel processing mode is activated (see Section~\ref{subsec:run-mode}).


\subsubsection{Post-processing and data outputs}
\label{subsec:post-processing}
Once the model is trained, for each input image, Mask R-CNN outputs a list with the detected objects (one binary mask array per object) with corresponding detection confidence score in range [0,1], and a classification label. We then implemented a series of post-processing steps to produce the final list of detections:
\begin{enumerate}
\item Select objects detected with a score above a configurable threshold (0.7 by default);
\item Merge overlapping or connected objects detected with the same classification label, or retain only the object with the largest score otherwise. 
\end{enumerate}
In Fig.~\ref{fig:sample-output} we show a typical output obtained after the post-processing stage on three different images, containing sample objects of all classes. For each image, we superimposed the detected objects with their masks (coloured by label) and the detection score.

\begin{figure}[!h]
\centering%
\includegraphics[scale=5.2]{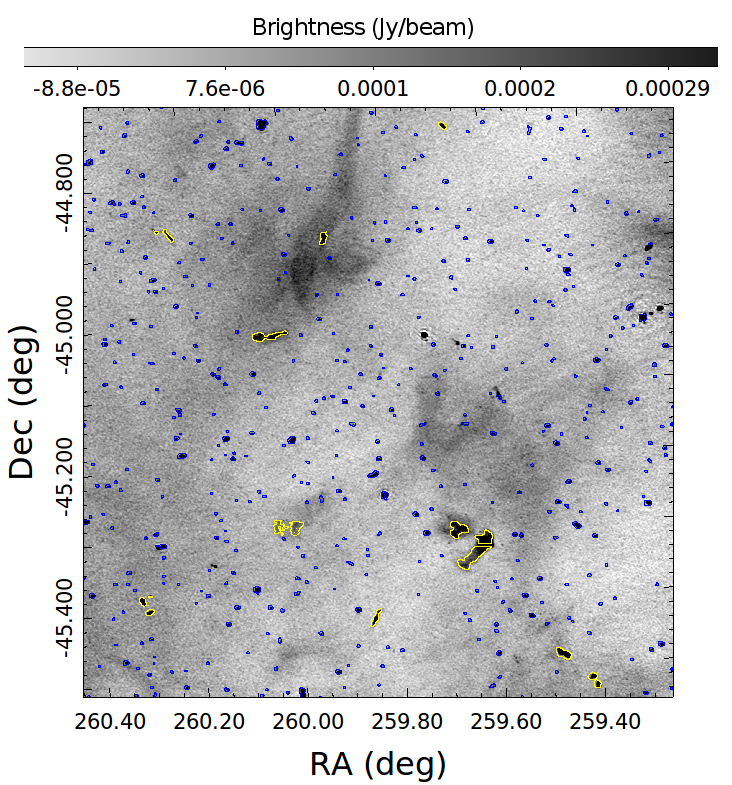}
\caption{Sample image (2000$\times$2000 pixels) taken from the \scorpio{} ASKAP survey with sources extracted by \toolname{} in parallel mode superimposed (blue: compact sources, yellow: extended).}
\label{fig:parallel-processing}
\end{figure}

\subsubsection{Run options and processing}
\label{subsec:run-mode}
\toolname{} can be run from the command line as:
\begin{codebkg}
python run.py [OPTIONS] [RUN MODE]
\end{codebkg}
Currently, three run modes \{\emph{train}, \emph{test}, \emph{detect}\} are supported. \emph{train} and \emph{test} modes are used to train (or re-train) a model and estimate detection and classification performance, respectively. In the \emph{detect} (or source finding) mode, the tool uses a trained model to detect objects from an input image, finally producing source catalogue outputs.\\
We considerably improved the original tool configuration, providing over 60 command-line arguments to specify the parameters to be used in training, testing, and detection mode. Table~\ref{tab:command-line-arguments} includes the list of all accepted command-line arguments, along with a description, default and accepted values.\\
To support detection on larger images, we produced a parallel implementation, based on \emph{mpi4py} library\footnote{\url{https://mpi4py.readthedocs.io/en/stable/}}, splitting the detection task on multiple image tiles across different processors, eventually merging individual detections found in adjacent tiles. The parallel version, still experimental at the time of writing, can be run as:
\begin{codebkg}
mpirun -np [NPROC] python run.py [OPTIONS] [RUN MODE]
\end{codebkg}
where \texttt{\textsc{nproc}} is the number of MPI processes to be used. Additional options are provided to configure the desired tile size and overlap. For example, in Fig.~\ref{fig:parallel-processing} we report the detections obtained in parallel mode (n$_{proc}$=4, tile size=512 pixels) over a 2000$\times$2000 pixel image taken from the \scorpio{} ASKAP survey.

\begin{table*}[]
\centering%
\scriptsize%
\caption{Object detection metrics (Completeness $C$, Reliability $R$) obtained with the best performing model over the test sample for different classes and telescope data sources. IoU and score threshold were set to 0.6 and 0.5, respectively. Metrics are not reported when the number of available true objects is smaller than 20.}
\label{tab:detection-performances}
\begin{tabular}{clcccccc}
\hline%
\hline%
\multirow{2}{*}{Metric} & \multirow{2}{*}{Telescope}  & \multicolumn{4}{c}{\footnotesize{\textsc{source}}} & & \\%
\cmidrule(lr){3-6}%
& & \footnotesize{\textsc{all}} & \footnotesize{\textsc{compact}} & \footnotesize{\textsc{extended}} & \footnotesize{\textsc{extended-multisland}} & \footnotesize{\textsc{spurious}} & \footnotesize{\textsc{flagged}}  \\%
\hline%
\multirow{4}{*}{C (\%)} & \footnotesize{\textsc{vla}} &  78.5 & 82.2 & 74.2 & 68.3 & $-$ & $-$ \\%
& \footnotesize{\textsc{askap}} &  92.4 & 92.9 & 84.8 & $-$ & 47.9 & 79.7 \\%
& \footnotesize{\textsc{atca}} & 88.6 & 89.9 & 82.6 & $-$ & 31.8 & $-$ \\%
\cmidrule(lr){2-8}%
& \footnotesize{\textsc{all}} & 87.8 & 89.9 & 78.9 & 65.4 & 46.4 & 77.9 \\%
\hline%
\multirow{4}{*}{R (\%)} & \footnotesize{\textsc{vla}} & 48.5 & 40.8 & 87.2 & 88.1 & $-$ & $-$ \\%
& \footnotesize{\textsc{askap}} & 70.8 & 70.6 &  77.6 &  $-$ & 39.1 & 88.7 \\%
& \footnotesize{\textsc{atca}} & 67.7 & 66.5 & 85.0 & $-$ &  18.5 & $-$ \\%
\cmidrule(lr){2-8}%
& \footnotesize{\textsc{all}} & 63.5 & 61.5 & 84.4 &  87.6 & 36.7 & 88.0 \\%
\hline%
\end{tabular}
\end{table*}

\subsubsection{Data outputs}
\label{subsec:data-outputs}
The main data products produced in the \emph{train}/\emph{test} run mode are the trained model weights in HDF5 format, diagnostic images showing the obtained detections over the input images, and desired metric table files. 

In the \emph{detect} (or source finding) mode, \toolname{} currently produces these outputs for an input image:
\begin{enumerate}
\item a catalog of detected objects in json format\footnote{This output corresponds to the source island catalogue produced by traditional source finders.}, including these parameters:
\begin{itemize}
\item source position parameters: centroid, bounding box, pixel list, contour;
\item source flux parameters: integrated flux density, peak brightness, pixel brightness stats (e.g. min/max, median, standard deviation, etc), and flags (e.g. source at the map border, etc);
\item classification parameters: class label, classification score;
\end{itemize}
\item a DS9 region file, with detected sources (polygon region format), with classification tags applied;
\item a diagnostic plot (in PNG format), showing the input image and the detected sources, with superimposed pixel masks, bounding boxes, and classification scores;
\end{enumerate}
Source components for each detected island are deliberately not provided (e.g. no deblending or gaussian fitting is performed), as these can be obtained using standard source finders that already implement such processing stages, like \caesar{} or others widely used in the radio domain. Integration of \toolname{} and \caesar{} is in fact ongoing.

\section{Results}
\label{sec:results}

\subsection{Model training and parameter optimization}
\label{subsec:training}
The full image set described in Section~\ref{sec:dataset-training} was randomly split into a train, validation, and test set, containing 60\%, 10\%, and 30\% of the original sample size, respectively. Several training runs were then carried out on the train and validation sets to tune Mask R-CNN hyper-parameters, and understand their impact on the source detection and classification performance (see Sections~\ref{subsec:detection-performance} and \ref{subsec:classification-performance}). Most parameters were kept to their defaults (see~\ref{tab:appendix-1-default-parameters}, column 2), while we varied the following ones for our source detection purposes (see~\ref{tab:appendix-1-default-parameters}, column 3):
\begin{itemize}
\item \texttt{IMAGE\_MAX\_DIM}: Image size (in pixels) used to resize original images before being given as backbone network inputs. We set this parameter to an intermediate value (256) between the original image size (132) and the default (1024), to slightly increase the sky field of view "seen" by the model in full-image/mosaic applications, and to limit training runtimes (largely dependent on input image size);
\item \texttt{RPN\_ANCHOR\_SCALES}: Anchor scales control the typical object size in pixels Mask R-CNN is sensitive to. For a general-purpose model, aiming to detect both compact and extended sources, we have considered these scales\footnote{The network architecture supports up to five scales.} on the basis of expected source sizes in our dataset and input image resize choice: (8, 16, 32, 64, 128);
\item \texttt{RPN\_ANCHOR\_RATIOS}: Anchor ratios control the typical object aspect ratio Mask R-CNN is sensitive to. A value of 1 represents a square anchor, while values <1 or >1 represents elongated anchors along both dimensions. On the basis of expected aspect ratios in our dataset, to improve detection of single- or multi-island extended sources, we considered a wider set of anchor ratios: (0.2, 0.3, 0.5, 1.0, 2.0, 3.0, 4.0, 5.0);
\item \texttt{LOSS\_WEIGHTS}: By default all loss components are equally weighted before being summed up to compute the total model loss (see Section~\ref{subsec:loss-calculation}). In all runs, however, we noticed a large unbalance between segmentation loss (`rpn\_bbox\_loss', `mrcnn\_bbox\_loss', and `mrcnn\_mask\_loss') and classification loss (`rpn\_class\_loss' and the `mrcnn\_class\_loss') components, the former being an order of magnitude larger. The first three of these losses refer to how well the model is drawing bounding boxes and pixel masks over the objects, whereas the last two refer to how well the model is classifying the detected objects (higher loss implies poorer performance). This could be an indication that the model is learning more to segment sources than to classify them. We therefore downscaled classification losses by a factor 10 (e.g. we have set `rpn\_class\_loss' and the `mrcnn\_class\_loss' weights to 0.1), to rebalance all loss contributions. This led to a net increase in performance over the alternative model trained with equal loss weights.
\end{itemize}
The \textit{ResNet-101} backbone network was trained from scratch, as we obtained slightly superior performances in this case, compared to a pre-trained backbone on ImageNet dataset\footnote{\url{https://www.image-net.org/}}. A number of 150 epochs was found a suitable compromise to limit training runtimes and data overfitting (evaluated on validation set). Measured runtimes over train and validation sets on a DELL PowerEdge R740 server (2$\times$Intel Xeon Gold 6248R 3.0 GHz, 24 cores each, 512 GB RAM) equipped with 1 Quadro RTX 6000 GPU are of the order of $\sim$1800 s per epoch ($\sim$3 days to reach 150 epochs).
Once trained, however, the detection run is considerably faster, e.g. 2-3 s per image on a standard laptop without GPU (e.g. a quad-core Intel 2.3 GHz, 16 GB RAM).

\subsection{Source detection performance}
\label{subsec:detection-performance}
\begin{figure}[!h]
\centering%
\includegraphics[scale=0.4]{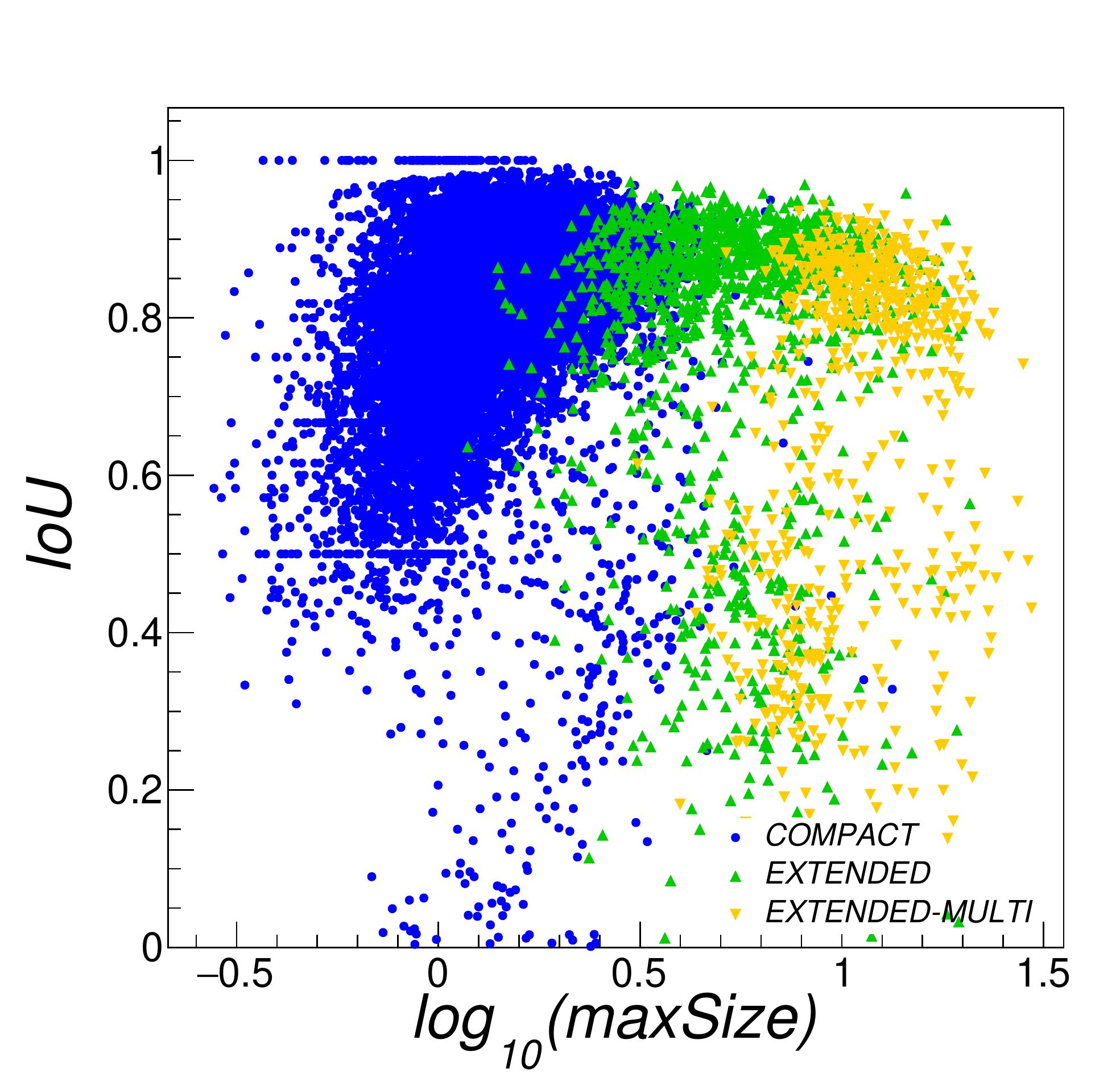}%
\caption{%
Distribution of Intersection-over-Union (IoU) values, computed between matched ground truth and detected object masks, for different true source classes as a function of the source max size.
}%
\label{fig:detection-ious}
\end{figure}

To evaluate how well the model detects radio sources (i.e. objects of class \texttt{\textsc{source}}, see Section~\ref{subsec:class-label-schema}), irrespective of their morphology classification, we considered the following conventional metrics, computed over the test set:
\begin{itemize}
\item \emph{Completeness} (C): Fraction of true sources matching a Mask R-CNN detected object of class label \texttt{\textsc{source}} and score larger than a specified threshold, i.e. the completeness for compact sources is the fraction of those sources that have been labelled as compact that get detected by the algorithm (above some threshold), regardless of what source class the algorithm assigns them to;
\item \emph{Reliability} (R): Fraction of Mask R-CNN detected objects, classified as \texttt{\textsc{source}} with score larger than a specified threshold, that indeed match to true sources;
\end{itemize}
A match is assumed between a true and a detected object if the IoU computed between their segmentation masks is larger than a specified threshold, set to 0.6 by default. Score threshold was set to 0.5. In Fig.~\ref{fig:detection-ious} we report the IoU distribution of matched sources for different source classes as a function of source max size, showing that masks are reconstructed with good accuracies on average (<IoU>$\sim$0.8 for all classes) above the default IoU threshold.
\begin{figure}[!h]
\centering%
\includegraphics[scale=0.4]{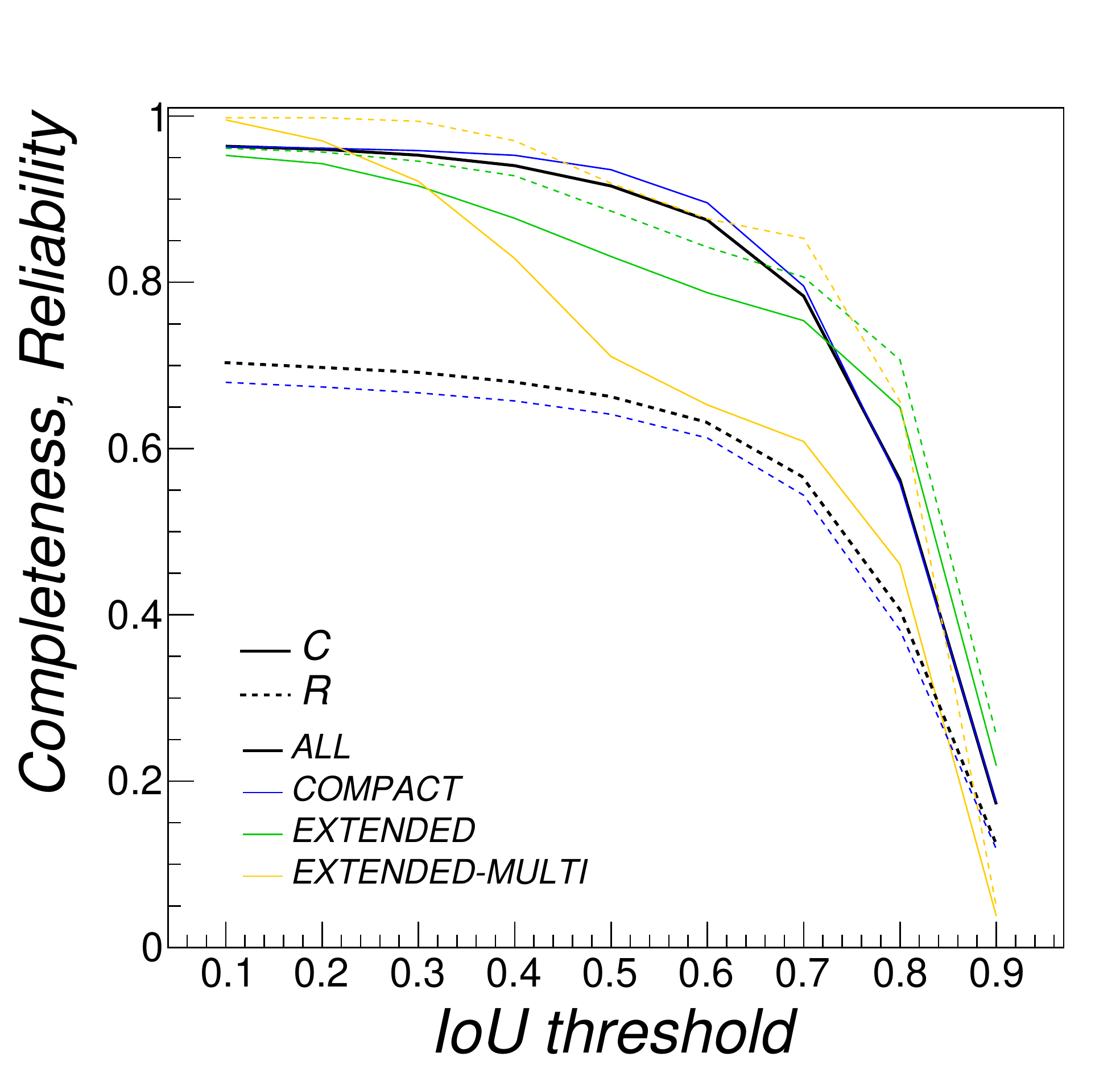}
\caption{%
Completeness (solid black line) and reliability (dashed black line) metrics as a function of the applied IoU match threshold. Results for compact, extended and extended-multisland sources are reported with blue, green and orange lines, respectively.
}%
\label{fig:det-metrics-vs-iouthr}
\end{figure}

Detection metrics obtained with the best performing model are reported in Table~\ref{tab:detection-performances}. As can be seen, compact sources are detected with high completeness ($\sim$90\%) and moderate reliability ($\sim$60\%), overall. Extended sources, particularly multi-island ones, are more frequently missed ($C$ ranging from 65\% to 80\%), but detected with much better reliabilities (>85\%). For the sake of completeness, we also reported metrics for spurious and flagged sources, although they are not relevant for source cataloguing scopes. Metrics were also re-computed by excluding sources located at the image borders, to search for possible boundary effects. We did not find a significant completeness increase overall ($\sim$2\%), pointing out that the model is indeed capable of detecting sources even if their segmentation mask is partially cut.
\begin{figure*}[!ht]
\centering%
\subtable[$C_{\tiny{\text{compact}}}$]{\includegraphics[scale=0.4]{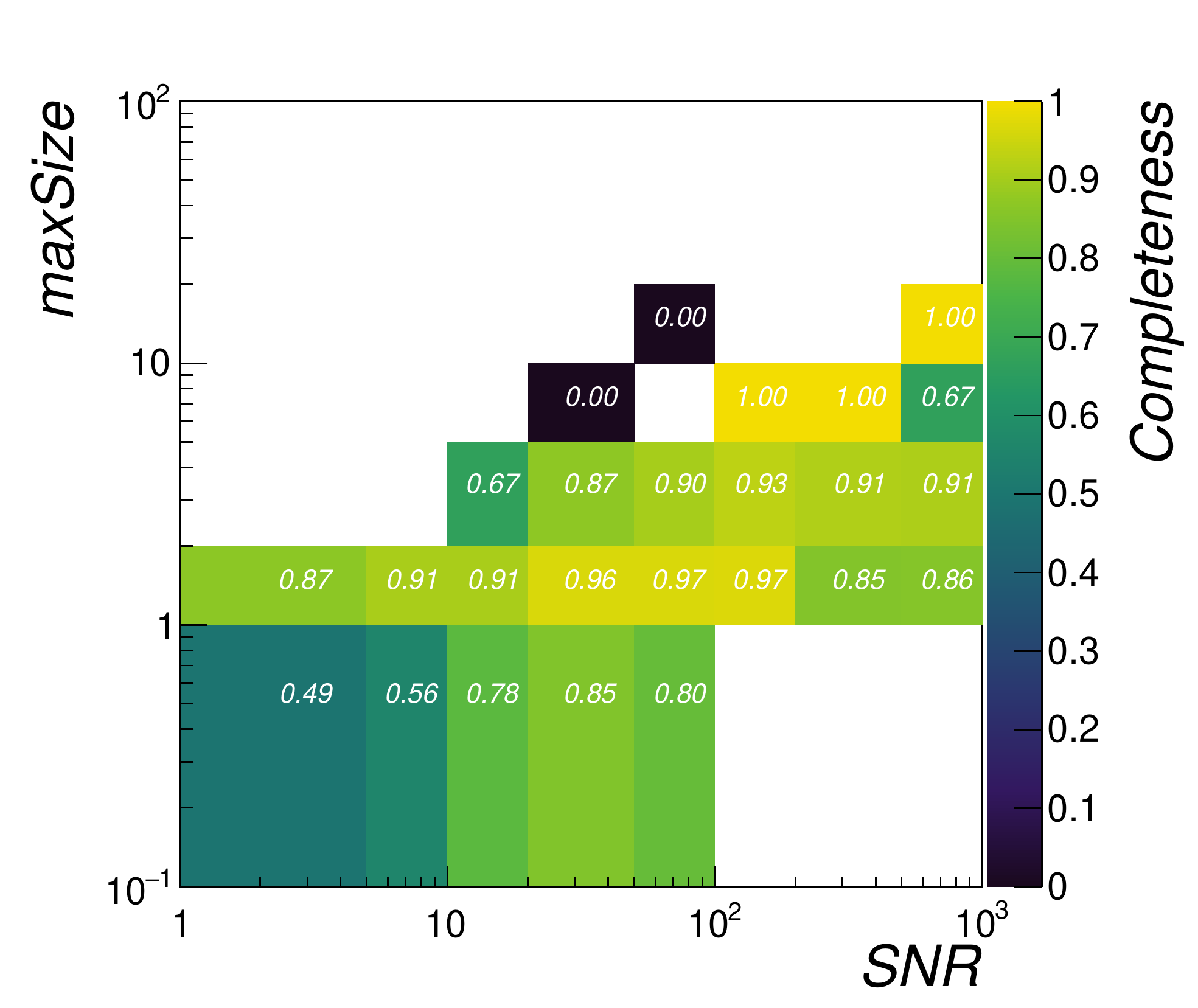}}%
\subtable[$R_{\tiny{\text{compact}}}$]{\includegraphics[scale=0.4]{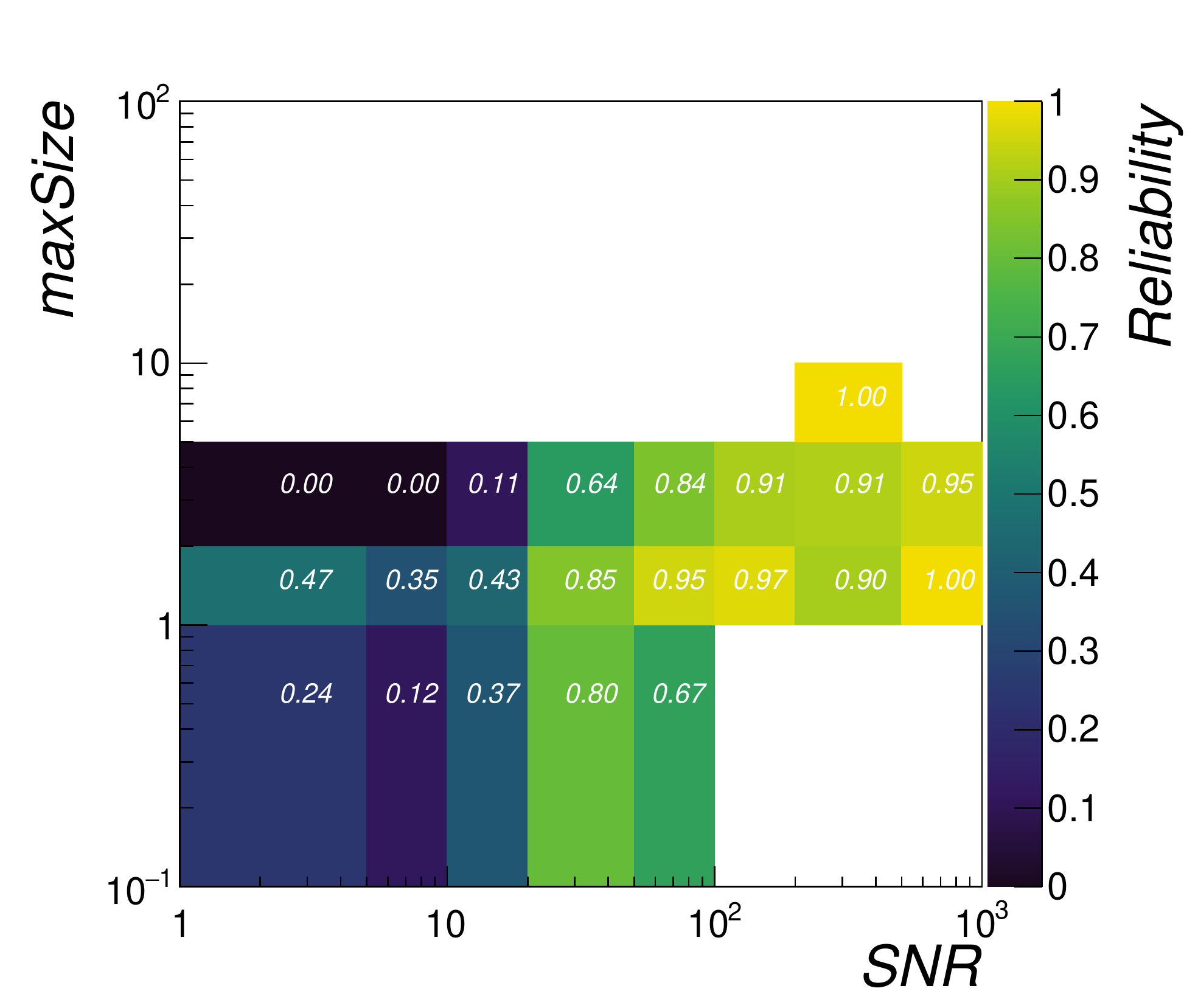}}%
\\%
\vspace{-0.42cm}
\subtable[$C_{\tiny{\text{extended+extended-multi}}}$]{\includegraphics[scale=0.4]{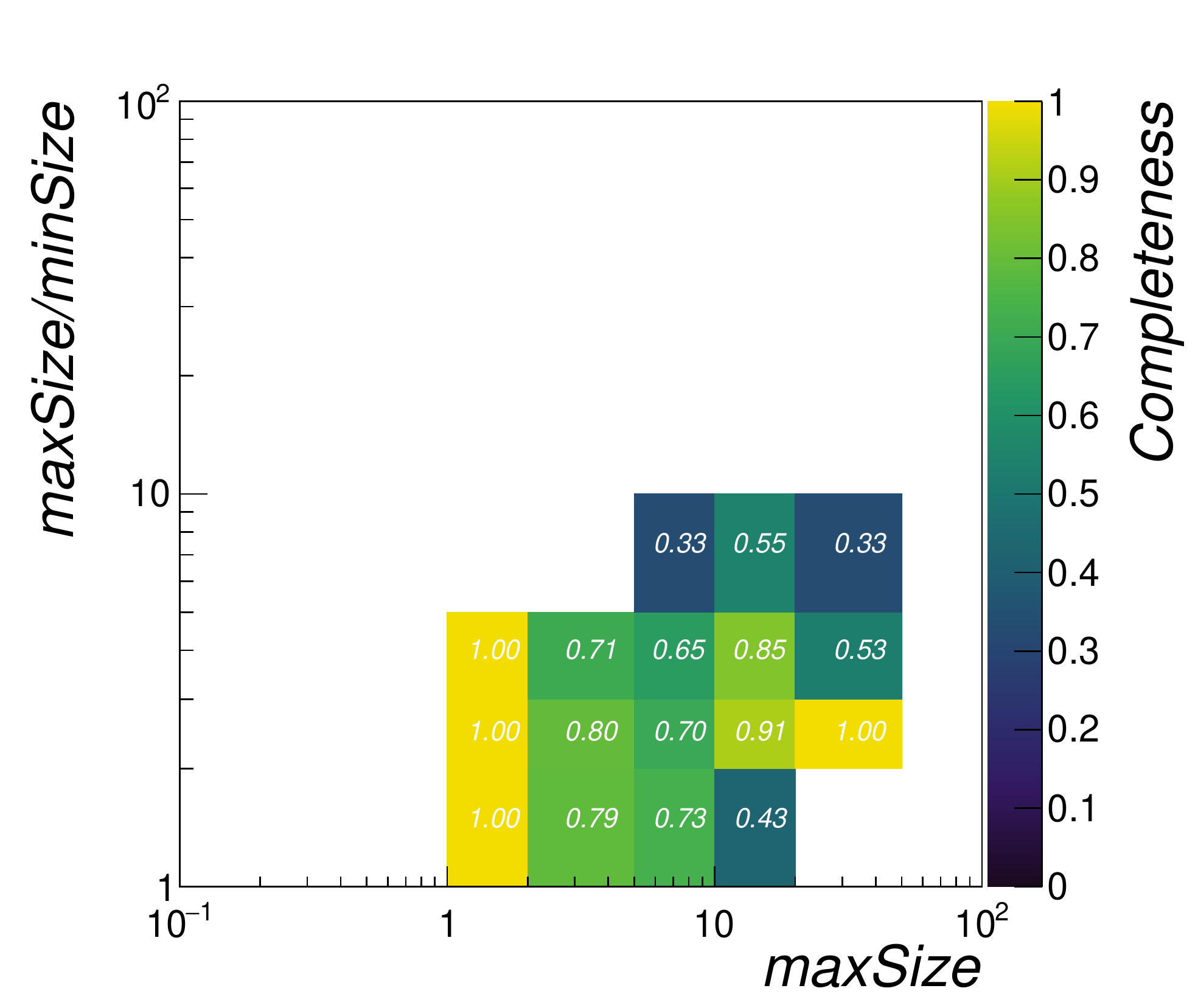}}%
\subtable[$R_{\tiny{\text{extended+extended-multi}}}$]{\includegraphics[scale=0.4]{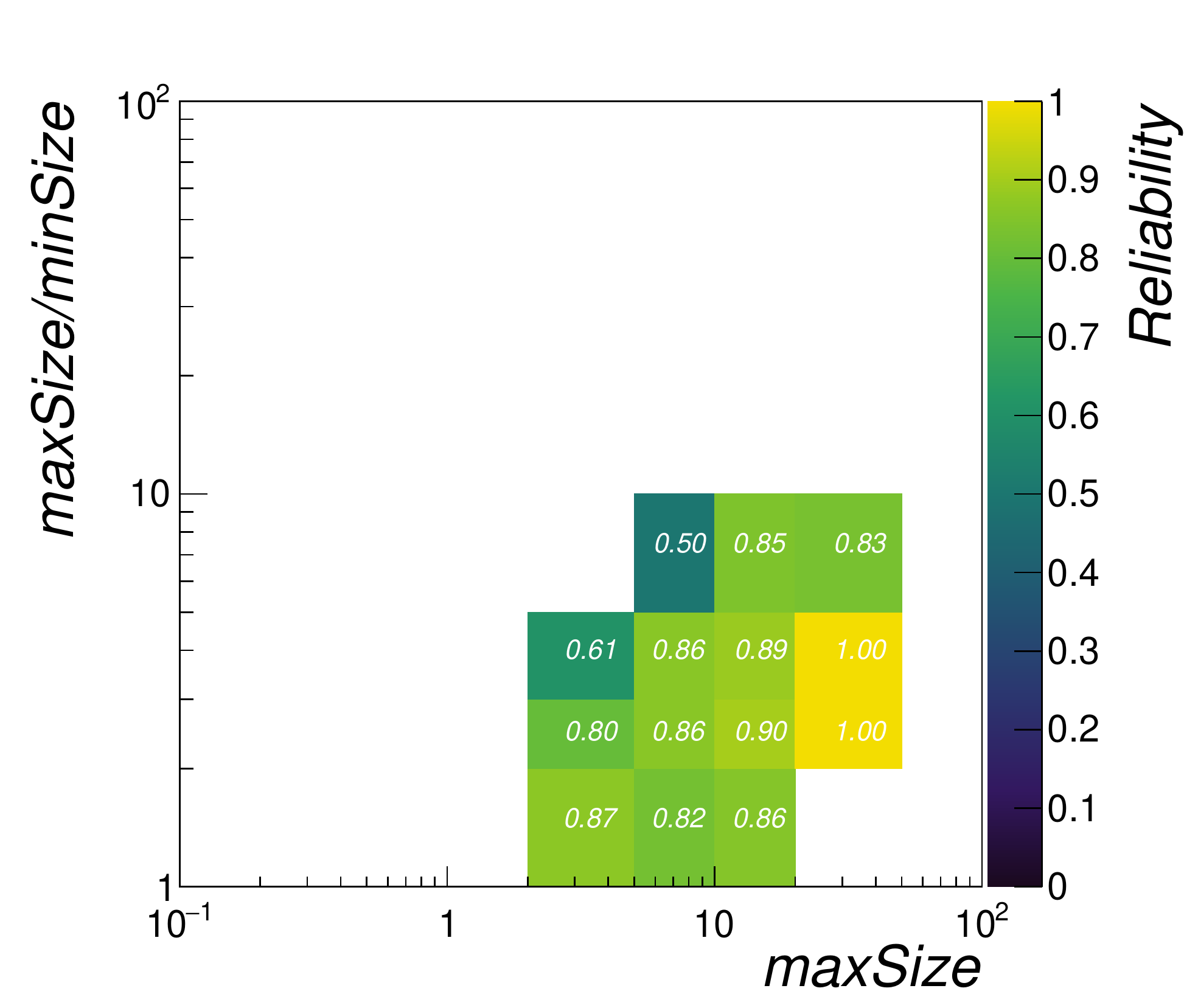}}%
\caption{%
Top: Completeness (left panel) and Reliability (right panel) metrics obtained over the test set on compact sources as a function of source SNR and max size. Bottom: Completeness (left panel) and Reliability metrics obtained over the test set on joint single- and multi-island extended sources as a function of source max size and aspect ratio.
}%
\label{fig:detection-perf}
\end{figure*}

In Fig.~\ref{fig:det-metrics-vs-iouthr} we report the detection metrics obtained as a function of the applied IoU match threshold for the entire data sample (black lines) and for each source class (colored lines). As expected, performances degrade as we increase the IoU threshold. The default value (0.6) was set to ensure that a reasonable fraction of the extended source mask is matched. We, however, observed from a visual inspection of the detections that this choice is suboptimal for compact sources, as it cuts good matches for sources at lower SNRs and smaller sizes compared to the beam (see Fig.~\ref{fig:detection-ious}), effectively decreasing the completeness. A lower threshold (0.4), increasing with the source size, would thus be preferable for matching compact sources to the fixed threshold default.

\begin{table*}[]
\centering%
\scriptsize%
\caption{Object detection metrics (Completeness $C$, Reliability $R$) obtained with the best performing model trained on VLA data images alone (see text) for different classes and telescope data sources. IoU and score threshold were set to 0.6 and 0.5, respectively. Metrics obtained on the VLA train data set are reported in rows (1) and (5), while results on the ASKAP/ATCA test sets are reported in rows (2-4) and (6-8).}
\label{tab:detection-performances-vlatrain}
\begin{tabular}{clcccc}
\hline%
\hline%
\multirow{2}{*}{Metric} & \multirow{2}{*}{Telescope}  & \multicolumn{4}{c}{\footnotesize{\textsc{source}}} \\%
\cmidrule(lr){3-6}%
& & \footnotesize{\textsc{all}} & \footnotesize{\textsc{compact}} & \footnotesize{\textsc{extended}} & \footnotesize{\textsc{extended-multisland}}  \\%
\hline%
\multirow{4}{*}{C (\%)} & \footnotesize{\textsc{vla}} & 81.9 & 82.2 & 83.7 & 78.9\\%
\cmidrule(lr){2-6}%
& \footnotesize{\textsc{askap+atca}} & 75.1 & 75.8 & 67.3 & 46.1 \\%
& \footnotesize{\textsc{askap}} & 75.6 & 76.2 & 64.1 & 60.0 \\%
& \footnotesize{\textsc{atca}} & 74.4 & 75.3 & 69.7 & 39.2 \\%
\hline%
\multirow{4}{*}{R (\%)} & \footnotesize{\textsc{vla}} & 54.5 & 45.4 & 92.5 & 83.9\\%
\cmidrule(lr){2-6}%
& \footnotesize{\textsc{askap+atca}} & 63.7 & 64.0 & 71.4 & 30.2 \\%
& \footnotesize{\textsc{askap}} & 67.8 & 69.2 & 55.1 & 24.9\\%
& \footnotesize{\textsc{atca}} & 58.1 & 56.4 & 85.0 & 37.1 \\%
\hline%
\end{tabular}
\end{table*}

\subsubsection{Performance for different radio surveys}
Our dataset contains images taken from a number of surveys, in which target objects are imaged at different resolution and with different background conditions. This could certainly help the model generalization capabilities for application to new survey data, provided that a reasonable balance in the number of objects from each reference survey is reached. This is unfortunately not yet the case for our dataset, nor for other available datasets (e.g. the RGZ, or similar ongoing projects). For this reason, it makes sense to inspect how the achieved performances depend on the considered survey. We therefore computed the model metrics on images from the three telescope source (VLA, ASKAP, ATCA), separately, reporting them in Table~\ref{tab:detection-performances}. 
In most cases, the number of available images is too small to draw firm conclusions, however, some interesting trends can be noticed where we have a sufficiently large statistics of objects. For example, VLA images are largely responsible for the low overall reliability obtained on compact sources. This is likely due to a larger fraction of noisy images compared to other survey data.
The reliability values found for extended sources among the available survey data are not easily comparable as they were found to have a different origin. In fact, a visual inspection of the false detections shows that the majority of them in VLA images are actually components of true multi-island sources that were not fully detected (e.g. only an extended component was detected, while the others went either undetected or misdetected as compact) and do not pass the IoU match threshold. On the other hand, in ASKAP and ATCA images (taken in the Galactic plane) they largely correspond to portions of extended sources or diffuse regions that were not labelled as true sources ($\sim$25\% of them located at the image borders), and to spurious or flagged extended sources in a minor percentage ($\sim$4\%).\\
To further characterize the impact of different surveys in the results, we trained a new Mask R-CNN model on VLA image data alone, and computed the performance metrics on the remaining test dataset made of exclusively ASKAP and ATCA images. Results, reported in Table~\ref{tab:detection-performances-vlatrain}, show a degradation of $\sim$10\% in model detection performance on test set with respect to previous analysis (Table~\ref{tab:detection-performances}), highlighting that using a mixture of survey data indeed improved the model generalization capabilities. In light of these indications, the preparation of curated datasets and transfer learning activities on a telescope- or survey-basis\footnote{To this aim, new crowdsourcing initiatives were launched within SKA precursors, such as the LOFAR and EMU Radio Galaxy Zoo.} are therefore unavoidable to port and fully exploit existing deep models (mostly trained on past survey or simulated data) on future SKA and precursor surveys.

\subsubsection{Performance against source parameters}
In Table~\ref{tab:detection-performances} we reported results over the full test set, but we expect detection performance to depend on object signal-to-noise (SNR) (particularly for compact sources) and shape (particularly for extended and spurious sources), so we computed metrics as a function of object SNR, size and aspect ratio. In Fig.~\ref{fig:detection-perf} (top panels) we report the completeness and reliability for compact sources as a function of source SNR and max size. Source counts obtained in each bin are reported in Fig.~\ref{fig:source-counts}. As expected, lower performances are obtained for fainter sources with size smaller than the synthesized beam size, although the observed reliability drop threshold is higher (SNR=10-20) compared to that observed in traditional finders on simulated data (SNR$\sim$5). From a visual inspection of the false detections, we noticed that many are dubious, e.g. they can well be real sources that were not annotated as ground truth, so there is a chance that the computed reliability is underestimated.\\In Fig.~\ref{fig:detection-perf} (bottom panels) we report the same metrics obtained on single- and multi-island extended sources as a function of the source max size and aspect ratio. Results show that very extended (>20 $\times$ synthesized beam size) and elongated sources are more easily missed by the model. As we did not observe any improvement with larger anchor scales (16, 32, 64, 128, 256), we conclude that this is likely due to the limited number of very extended sources in the training sample\footnote{At present, there are only 45 sources in the training sample with max size larger than 20 and aspect ratio larger than 3.}, and to the Mask R-CNN architecture which generally does not perform particularly well on thin and elongated shapes \citep{Looi2019}, "due to their thinness and curvature and therefore small area relative to the area of their associated ROIs" \citep{Frei2021}.

\begin{table*}[!ht]
\centering%
\scriptsize%
\caption{Object classification metrics (Recall $\mathcal{R}$, Precision $\mathcal{P}$, $F1$ score) obtained with the best performing model over the test sample for different classes and telescope data sources. Metrics are not reported when the number of available true objects is smaller than 20.}
\label{tab:precision-recall}
\begin{tabular}{clccccc}
\hline%
\hline%
Metric & Telescope & \footnotesize{\textsc{compact}} & \footnotesize{\textsc{extended}} & \footnotesize{\textsc{extended-multisland}} & \footnotesize{\textsc{spurious}} & \footnotesize{\textsc{flagged}}  \\%
\hline%
\multirow{4}{*}{$\mathcal{R}$ (\%)} & \footnotesize{\textsc{vla}} & 98.5 & 81.1 & 90.6 & $-$ & $-$ \\%
& \footnotesize{\textsc{askap}} & 99.5 & 70.2 & $-$ & 88.2 & 81.7 \\%
& \footnotesize{\textsc{atca}} & 99.5 & 81.0 & $-$ & 61.4 & $-$ \\%
\cmidrule(lr){2-7}%
& \footnotesize{\textsc{all}} & 99.3 & 78.5 & 88.2 & 85.7 & 80.5 \\%
\hline%
\multirow{4}{*}{$\mathcal{P}$ (\%)} & \footnotesize{\textsc{vla}} & 97.0 & 85.7 & 90.8 & $-$ & $-$ \\%
& \footnotesize{\textsc{askap}} & 97.7 & 89.5 & $-$ & 97.2 & 91.3 \\%
& \footnotesize{\textsc{atca}}& 97.7 & 93.4 & $-$ & 84.4 & $-$ \\%
\cmidrule(lr){2-7}%
& \footnotesize{\textsc{all}} & 97.6 & 88.8 & 89.3 & 96.3 & 90.5\\%
\hline%
\multirow{4}{*}{$F1$ (\%)} & \footnotesize{\textsc{vla}} & 97.7 & 83.4 & 90.7 & $-$ & $-$ \\%
& \footnotesize{\textsc{askap}} & 98.6 & 78.7 & $-$ & 92.5 & 86.2 \\%
& \footnotesize{\textsc{atca}} & 98.6 & 86.7 &  $-$ & 71.1 & $-$ \\%
\cmidrule(lr){2-7}%
& \footnotesize{\textsc{all}}  & 98.4 & 83.4 & 88.8 & 90.7 & 85.2 \\%
\hline%
\end{tabular}
\end{table*}

\subsection{Source classification performance}
\label{subsec:classification-performance}
\begin{figure}[!h]
\centering%
\includegraphics[scale=0.45]{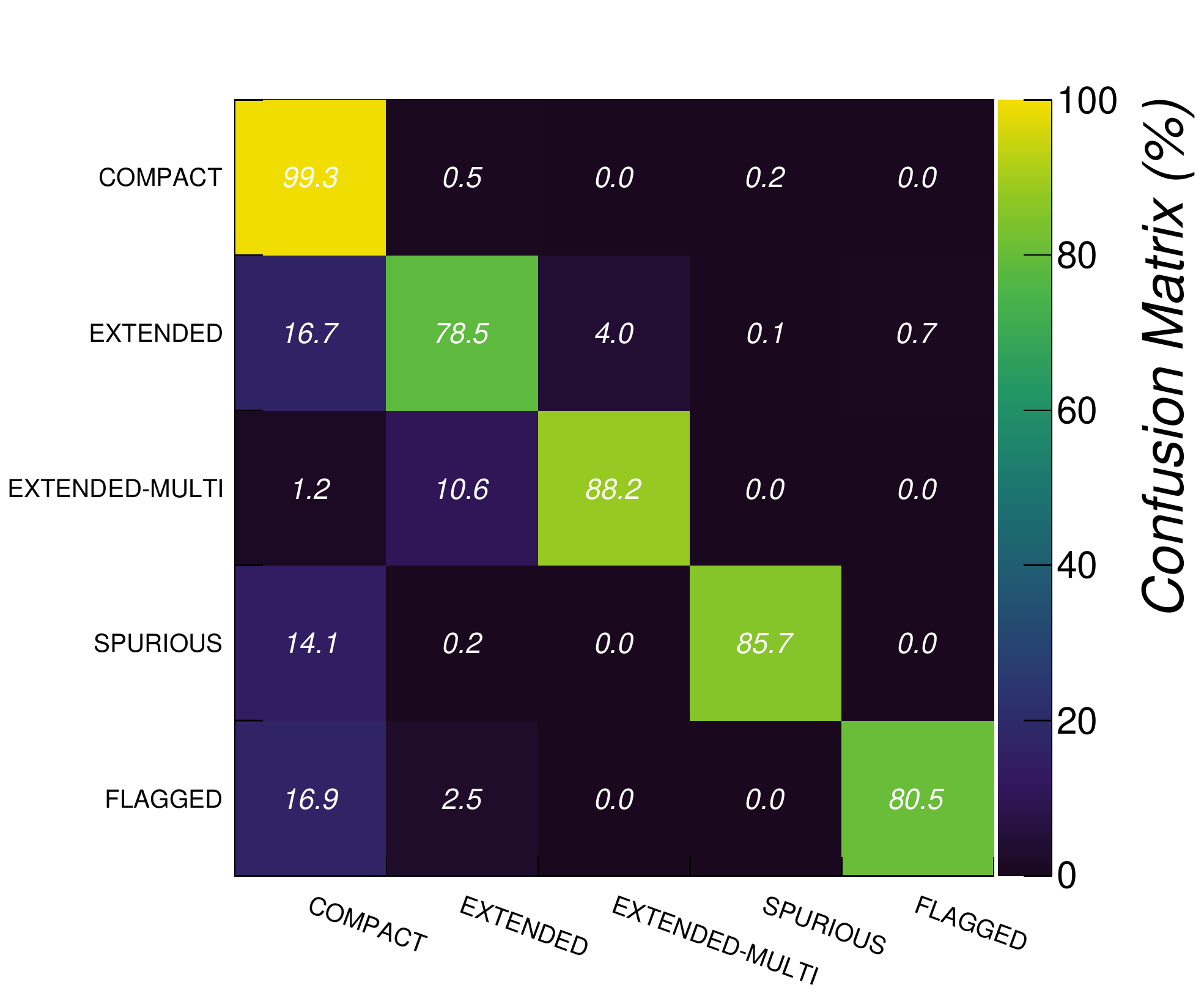}%
\caption{%
Confusion matrix obtained with the best performing model over the test set.
Rows (y-axis) are the true class labels, while columns (x-axis) are the predicted class labels.
}%
\label{fig:confusion-matrix}
\end{figure}

\begin{figure*}[!ht]
\centering%
\subtable[]{\includegraphics[scale=0.19]{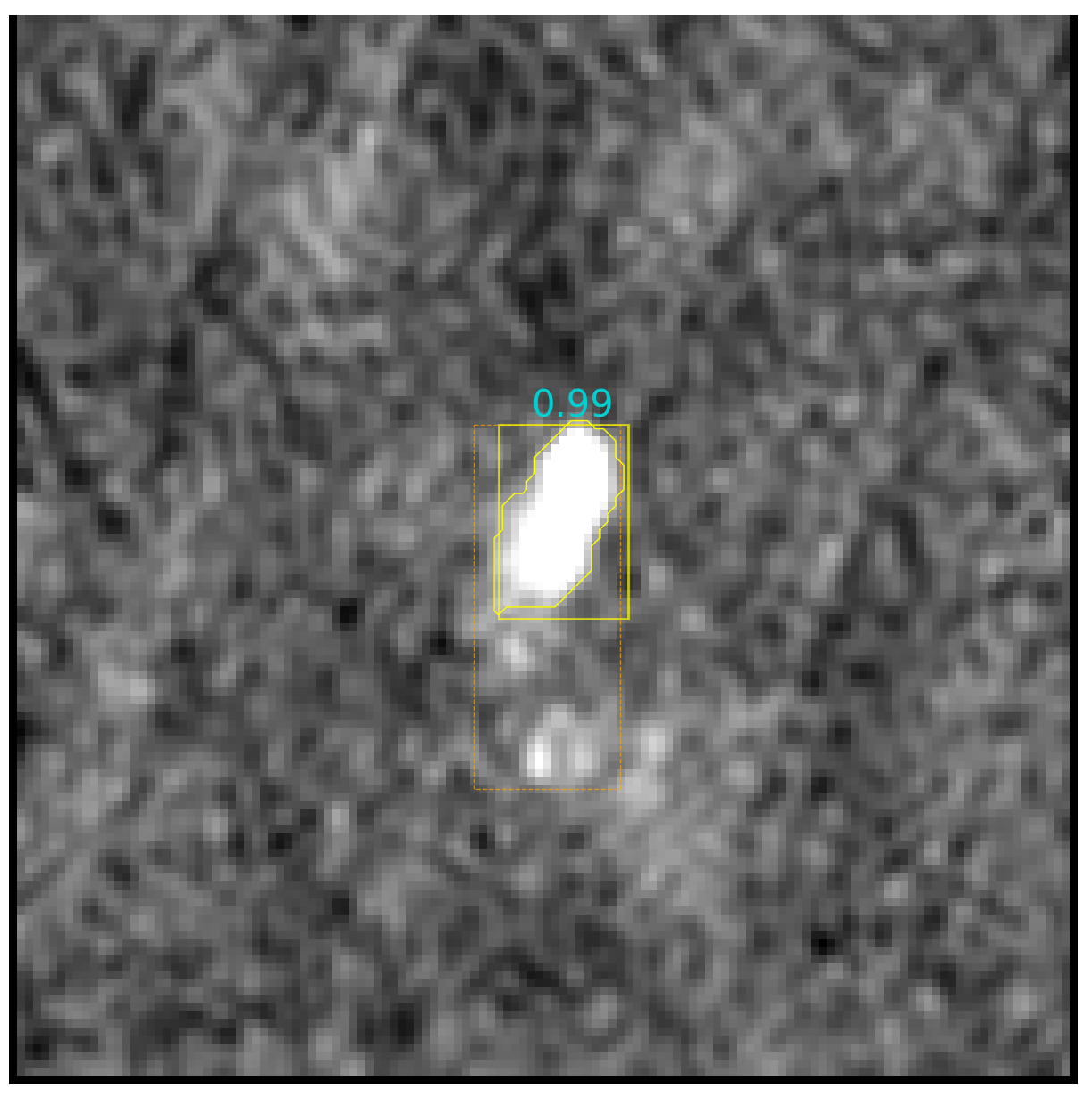}\label{fig:sample-misclassifications_1}}%
\subtable[]{\includegraphics[scale=0.19]{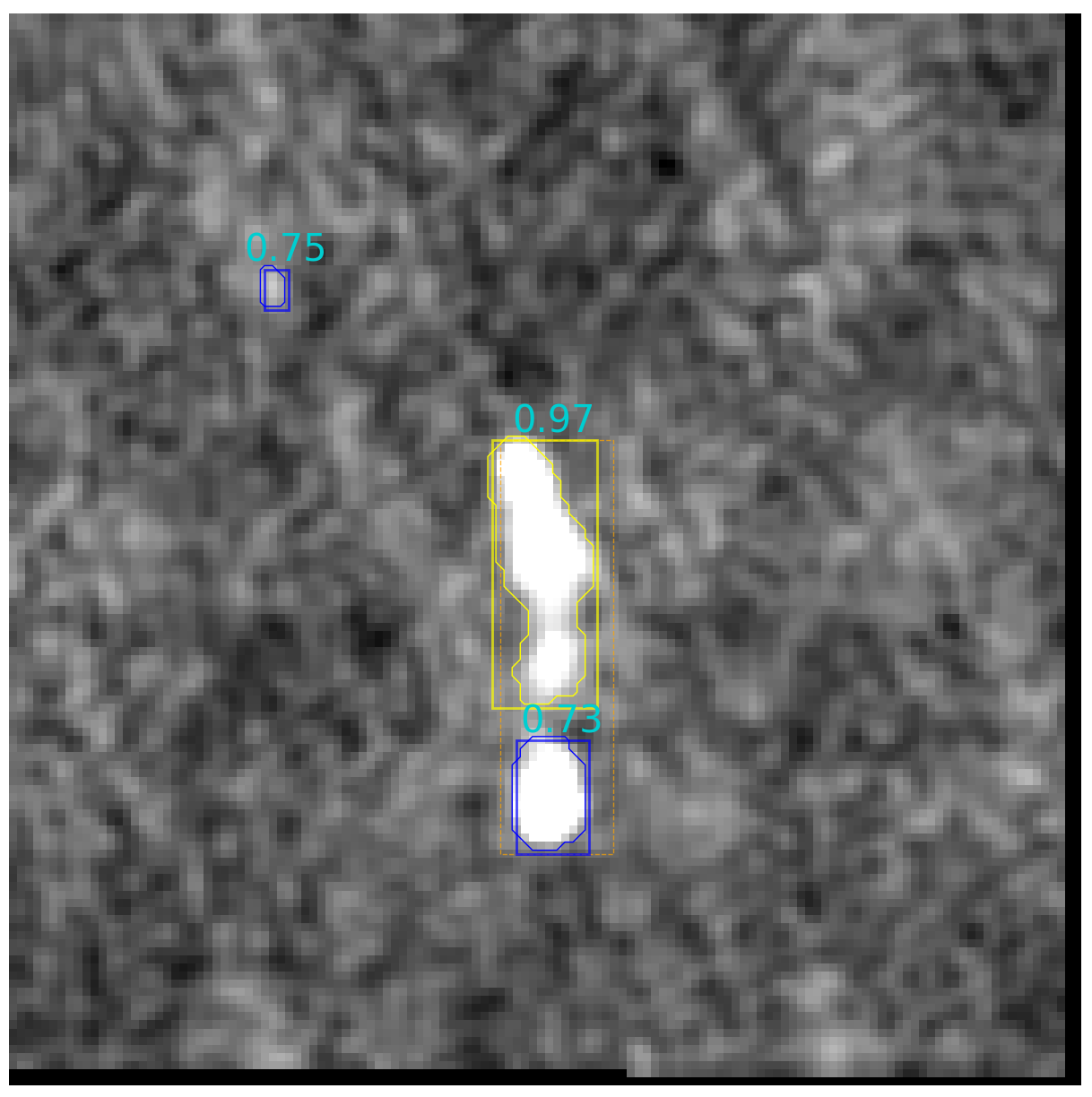}\label{fig:sample-misclassifications_2}}%
\subtable[]{\includegraphics[scale=0.19]{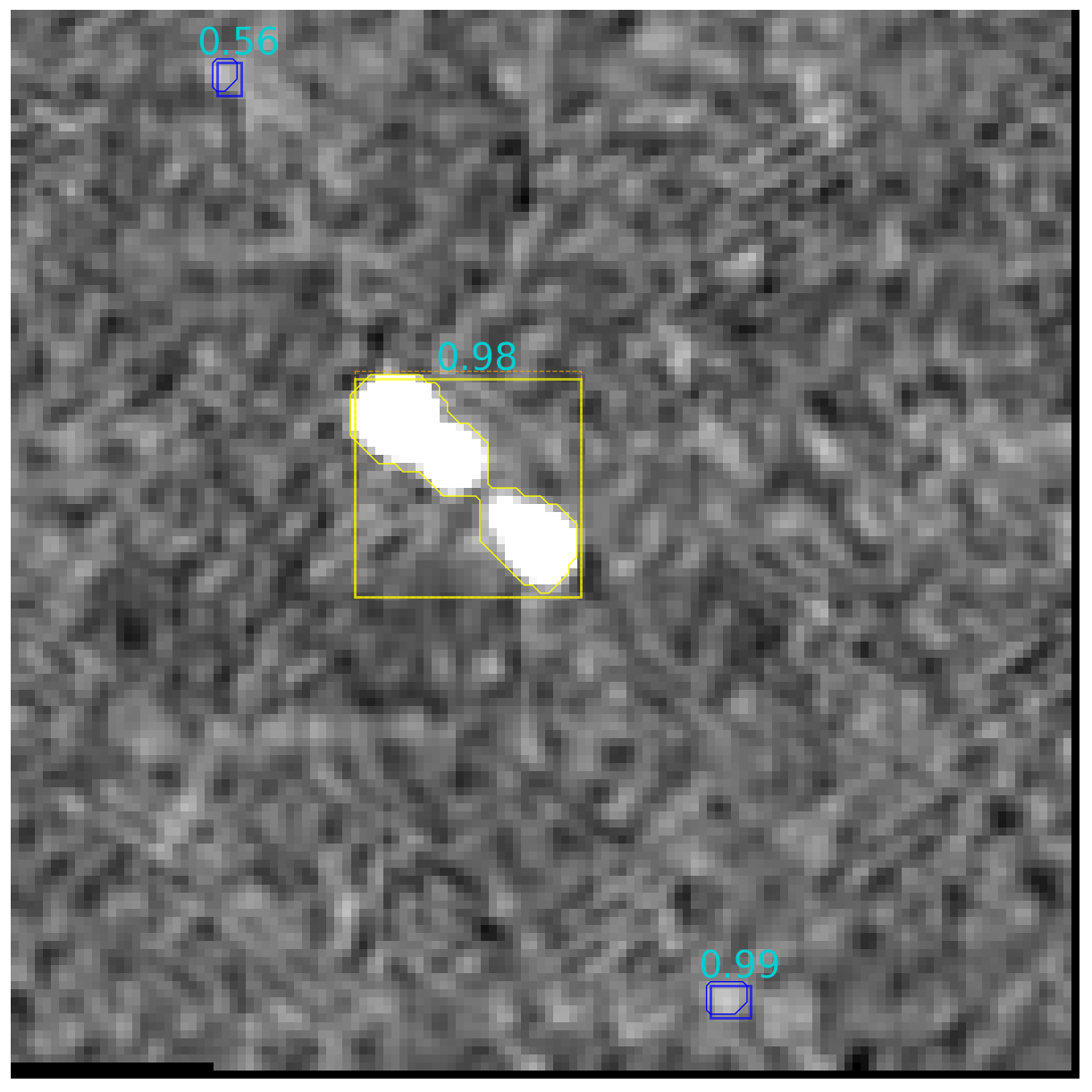}\label{fig:sample-misclassifications_3}}%
\caption{%
Sample images with misclassified multi-island extended sources.
}%
\label{fig:sample-misclassifications}
\end{figure*}

We computed the following conventional metrics on the test set to evaluate how well the model classifies the detected objects, matched to ground truth objects, into the defined classes:
\begin{itemize}
\item \emph{Recall} ($\mathcal{R}$): Fraction of true objects correctly classified by the model, out of the total number of true objects matching a Mask R-CNN detected object (as defined in Section~\ref{subsec:detection-performance});
\item \emph{Precision} ($\mathcal{P}$): Fraction of correctly classified objects out of the total number of detected objects, matching to a true object (as defined in Section~\ref{subsec:detection-performance});
\item \emph{F1 score}: the harmonic mean of precision and recall:
\begin{equation}
F1=2\times\frac{\mathcal{P}\times\mathcal{R}}{\mathcal{P}+\mathcal{R}}
\end{equation}
\end{itemize}
Classification metrics per each class and telescope data source are reported in Table~\ref{tab:precision-recall}. Overall, we obtained good classification performances ($F1$>80\%) for all classes, without significant differences among the three telescope survey data, except for extended sources in ASKAP which are more misclassified as compact sources compared to other surveys. Unfortunately, the number of available data is rather small in this case to confirm this trend.\\
In Fig.~\ref{fig:confusion-matrix} we report the confusion matrix obtained 
with the best performing model over the test set. Matrix elements $c_{ij}$ represent the fraction of true objects of class $i$ that are classified as class $j$. As can be seen, a large percentage ($>$80\%) of the detected sources (particularly compact sources) are correctly classified in their morphological class. 
About 17\% of the labelled extended sources are misclassified as compact sources, while $\sim$10\% of multi-island extended are classified as extended. From a visual inspection of the results, we noticed that 
all extended sources classified in the compact class are small and more roundish compared to the rest of extended sample, e.g. maxSize<5 and aspectRatio<3. Misclassifications of multi-island extended sources in the extended class are instead due to the fact that one or more source islands are either not detected (like the example shown in Fig.~\ref{fig:sample-misclassifications_1}) or not associated to the same object (e.g. classified as compact in most cases, like for example in Fig.~\ref{fig:sample-misclassifications_2}).
We also found cases (e.g. see Fig.~\ref{fig:sample-misclassifications_3}) in which the true source islands are disjoint, but very close to each other (1 or 2 pixels apart). Mask R-CNN was able to segment the entire object, but their detected masks were merged into one, leading to the final classification as extended source.\\It is interesting to observe the misclassification rate for spurious and flagged sources, as this eventually affect the source catalogue reliability. Roughly, 14\% of labelled spurious sources are classified as compact sources, while $\sim$20\% of flagged sources are classified as non-flagged. Although it is certainly desired to reduce the contamination percentage in the future, it is worth to stress that this is already a huge step forward towards catalogue automation, compared to the current scenario existing in most small- or large-area surveys, in which flagged and spurious sources are removed by visual inspection. 

\section{Summary and outlook}
\label{sec:summary}
In this work we have presented a new source finding tool, based on the Mask R-CNN instance segmentation framework, for detecting and classifying compact, extended, flagged and spurious sources in radio continuum maps. The method was tested on a dataset of images extracted from different radio surveys, including ASKAP EMU Early Science and pilot data, and performances were studied for different parameter values. The implemented tool offers these novelty aspects when compared to traditional radio source finders or other existing deep learning tools:
\begin{enumerate}
\item It provides object segmentation masks and not only bounding boxes as existing deep source finders;
\item It is capable of automatically detecting imaging artefacts around bright sources and flagged sources to be excluded from the source catalogue. Existing finders do not provide this feature;
\item It is capable of automatically detecting extended sources broken down into multiple islands. Traditional source finders lack this feature;
\item It is capable of running on large radio maps in parallel mode, not just on small cutouts as in many traditional and deep learning finders;
\item It makes use and was tested on real data from different radio telescopes, although with limited and unbalanced samples, compared to most tools that were trained on simulated data or real data from one specific telescope.
\end{enumerate}
There are also features that our tool intentionally does not provide with respect to traditional finders, e.g. component detection and measurement through island fitting, as the final goal is making existing and new tools interoperable, exploiting their offered benefits and peculiarities.\\
Overall, we found promising source detection performance metrics. On compact sources, we obtained a good completeness ($\sim$90\%), but a modest reliability ($\sim$60\%), mostly driven by the poor estimate on VLA data. Both metrics are not yet at the same level of those reported by traditional finders, which are however in most cases estimated on simulated data. We therefore conclude that at the present state, traditional source finders are still to be preferred when searching for compact and point-like sources.\\Performances on extended sources ($\sim$80\%) are inferior with respect to compact sources, particularly on multi-islands. As traditional finders do not report metrics for real single-island extended sources (only on ideally-modelled simulated sources, e.g. see \citealt{Riggi2019} or \citealt{Hopkins2015}) nor they detect multi-island extended sources as a unique object, it is not possible to make a qualitative comparison with our results. They are a respectable starting point, however, given the current scenario, that we expect to largely improve by adding more extended source data.\\
After the detection step, we obtained good classification performances (metrics above 80\%) for all classes of sources. Unfortunately, the performances achieved on spurious and flagged sources are suboptimal compared to the other two classes. This is in large part due to the limited number of available and labelled data for them. Nevertheless, this is already a significant improvement over the current approach followed in many surveys, entirely based on a manual detection.\\
The obtained results motivate a number of activities to be done in the very near future to tackle some of the encountered limitations (e.g. a lack of labelled data for some classes) that negatively affect the model performance. We are in fact currently increasing the size of our dataset with additional radio survey data, from ASKAP and other SKA precursors, and also generating synthetic data through generative adversarial networks (GANs). The latter activity is crucial as the efforts spent in visual inspection and source labelling of new data have become unsustainable.\\Another plan to address the poorer performance on extended objects is to instead consider alternative deep model frameworks, trained on the same dataset. For example, Rotated Mask R-CNN\footnote{\url{https://github.com/mrlooi/rotated_maskrcnn}}, claims to achieve better performance on elongated structures \citep{Looi2019,Frei2021}, and could be adapted for radio source detection scopes. We are also currently surveying alternative object detection frameworks on the same dataset \citep{Sortino2022}. Some of the tested models, for instance the \emph{Tiramisu} model, were reported to achieve significantly better performance on imaging artefacts \citep{Pino2021}.\\On a longer-term, we also plan to explore the possibility of detecting other classes of sources, e.g. diffuse sources or filaments.

Finally, we are working to allow \toolname{} to work not only as a standalone source finder (as described in this work), but also in combination with traditional tools, for example as a classifier stage applied to existing source finders catalogue outputs. Such a tool would certainly be valuable not only for the radio astronomy domain, but also for the astronomical community working in other fields. For this reason, we plan to integrate it as a service in the future European Open Science Cloud (EOSC) infrastructure. In this context, the NEANIAS \citep{Sciacca2020} and the CIRASA \citep{CIRASA} projects are building prototype solutions for astronomical source finding and visualization on the cloud, scalable to larger infrastructures, such as those expected to be deployed in the SKA Regional Centres. The Mask R-CNN detector described in this work is already integrated as a supported application within the developed space services\footnote{\url{https://github.com/SKA-INAF/caesar-rest}} (more details reported in \citealt{CIRASA}), although its usage is at present limited to small user images and cutouts, with plans to extend the integration to the full pipeline.

\section*{Acknowledgements}
\small{
The Australian SKA Pathfinder is part of the Australia Telescope National Facility which is managed by CSIRO. Operation of ASKAP is funded by the Australian Government with support from the National Collaborative Research Infrastructure Strategy. Establishment of the Murchison Radio-astronomy Observatory was funded by the Australian Government and the Government of Western Australia. This work was supported by resources provided by the Pawsey Supercomputing Centre with funding from the Australian Government and the Government of Western Australia. We acknowledge the Wajarri Yamatji people as the traditional owners of the Observatory site.

Part of the research leading to these results has received funding from the INAF PRIN TEC programme (CIRASA) and the European Commissions Horizon 2020 research and innovation programme under the grant agreement No. 863448 (NEANIAS).

We thank the authors of the following software tools and libraries that have been extensively used in this work: \caesar{}~\citep{Riggi2016,Riggi2019}, astropy~\citep{astropy2013,astropy2018}, ds9~\citep{Joye2003}, \textsc{dvc}.
}

\section*{Data Availability}
\small{
The data products used in this work will be made publicly available once the full \toolname{} pipeline is released and data sharing agreements are established (mainly for ASKAP and other private observatory data included in the dataset).\\All the code written for \toolname{} is also publicly available on the GitHub repository \url{https://github.com/SKA-INAF/caesar-mrcnn/}, under the GNU General Public License v3.0\footnote{\url{https://www.gnu.org/licenses/gpl-3.0.html}}. The weights files for the trained models are available on Zenodo at \url{https://doi.org/10.5281/zenodo.7377723}.
}

\appendix%
\onecolumn%

\section{Configuration Parameters}

\setcounter{table}{0}


\begin{flushleft}
\captionsetup{labelfont=bf}
\captionof{table}{
Default Mask R-CNN hyperparameters and the values used for \toolname{}. Default values are taken from: \url{https://github.com/matterport/Mask\_RCNN/blob/master/mrcnn/config.py}.}%
\label{tab:appendix-1-default-parameters}
\footnotesize{%
\begin{threeparttable}
\begin{tabular}{lll}
\hline%
\hline%
\textbf{Parameter} & \textbf{Default} & \toolname{} \\ \hline
\texttt{BACKBONE} & resnet101 & resnet101 \\
\texttt{COMPUTE\_BACKBONE\_SHAPE} & None & Default \\
\texttt{BACKBONE\_STRIDES} & {[}4, 8, 16, 32, 64{]} & {[}4, 8, 16, 32, 64{]} \\
\texttt{FPN\_CLASSIF\_FC\_LAYERS\_SIZE} & 1024 & Default \\
\texttt{TOP\_DOWN\_PYRAMID\_SIZE} & 256 & Default \\
\texttt{RPN\_ANCHOR\_SCALES} & (32, 64, 128, 256, 512) & (8, 16, 32, 64, 128) \\
\texttt{RPN\_ANCHOR\_RATIOS} & {[}0.5, 1, 2{]} & {[}0.2, 0.3, 0.5, 1.0, 2.0, 3.0, 4.0, 5.0{]} \\
\texttt{RPN\_ANCHOR\_STRIDE} & 1 & Default \\
\texttt{RPN\_NMS\_THRESHOLD} & 0.7 & Default \\
\texttt{RPN\_TRAIN\_ANCHORS\_PER\_IMAGE} & 256 & 256 \\
\texttt{PRE\_NMS\_LIMIT} & 6000 & Default \\
\texttt{POST\_NMS\_ROIS\_TRAINING} & 2000 & Default \\
\texttt{POST\_NMS\_ROIS\_INFERENCE} & 1000 & Default \\
\texttt{USE\_MINI\_MASK} & True & False \\
\texttt{MINI\_MASK\_SHAPE} & (56, 56) & N/A \\
\texttt{IMAGE\_RESIZE\_MODE} & square & square \\
\texttt{IMAGE\_MIN\_DIM} & 800 & 256 \\
\texttt{IMAGE\_MAX\_DIM} & 1024 & 256 \\
\texttt{IMAGE\_MIN\_SCALE} & 0 & Default \\
\texttt{IMAGE\_CHANNEL\_COUNT} & 3 & Default \\
\texttt{MEAN\_PIXEL} & {[}123.7, 116.8, 103.9{]} & {[}0,0,0{]} \\
\texttt{TRAIN\_ROIS\_PER\_IMAGE} & 200 & 256 \\
\texttt{ROI\_POSITIVE\_RATIO} & 0.33 & Default \\
\texttt{POOL\_SIZE} & 7 & Default \\
\texttt{MASK\_POOL\_SIZE} & 14 & Default \\
\texttt{MASK\_SHAPE} & {[}28, 28{]} & Default \\
\texttt{MAX\_GT\_INSTANCES} & 100 & 100 \\
\texttt{RPN\_BBOX\_STD\_DEV} & {[}0.1, 0.1, 0.2, 0.2{]} & Default \\
\texttt{BBOX\_STD\_DEV} & {[}0.1, 0.1, 0.2, 0.2{]} & Default \\
\texttt{DETECTION\_MAX\_INSTANCES} & 100 & Default \\
\texttt{DETECTION\_MIN\_CONFIDENCE} & 0.7 & 0 \\
\texttt{DETECTION\_NMS\_THRESHOLD} & 0.3 & 0.3 \\
\texttt{LEARNING\_RATE} & 0.001 & 0.0005 \\
\texttt{OPTIMIZER} & SGD & ADAM \\
\texttt{LEARNING\_MOMENTUM} & 0.9 & N/A \\
\texttt{WEIGHT\_DECAY} & 0.0001 & Default \\
\texttt{LOSS\_WEIGHTS} & rpn\_class\_loss: 1.0, & rpn\_class\_loss: 1.0,\\
& rpn\_bbox\_loss: 1.0, & rpn\_bbox\_loss: 0.1, \\%
& mrcnn\_class\_loss: 1.0, & mrcnn\_class\_loss: 1.0,\\%
& mrcnn\_bbox\_loss: 1.0, & mrcnn\_bbox\_loss: 0.1,\\%
& mrcnn\_mask\_loss: 1.0 & mrcnn\_mask\_loss: 0.1 \\%
\texttt{USE\_RPN\_ROIS} & True & Default \\
\texttt{TRAIN\_BN} & False & Default \\
\texttt{GRADIENT\_CLIP\_NORM} & 5.0 & Default \\ \hline
\end{tabular}
\end{threeparttable}
}
\end{flushleft}

{\footnotesize
\begin{longtable}[c]{llll}
\caption{List of \toolname{} command line options, including a short description, and accepted values.}
\label{tab:command-line-arguments}\\%
\hline%
\textbf{Option} & \textbf{Description} & \textbf{Default Value} & \textbf{Accepted Values} \\%
\hline%
\endfirsthead%
\multicolumn{4}{c}%
{{\bfseries Table \thetable\ continued from previous page}} \\
\endhead
\hline
\endfoot
\endlastfoot
\rowcolor{lightgray}%
\multicolumn{4}{c}{\footnotesize{REQUIRED OPTIONS}} \\%
\hline%
\textless{}command\textgreater{} & Indicates whether the model should train on the training set, & required & train/test/detect\\%
& evaluate performance of a trained model on the test set & & \\%
& or use a trained model to run detection on a provided image & &\\%
\hline%
\rowcolor{lightgray}%
\multicolumn{4}{c}{DATA LOADING \& PRE-PROCESSING} \\%
\hline%
\emph{-{}-imgsize} & Size the input image is resized to & 256 & Integers \\%
\emph{-{}-grayimg} & Disable image conversion to RGB & False & N/A \\%
\emph{-{}-no\_uint8} & Disable image pixel value conversion to uint8 & False & N/A \\%
\emph{-{}-no\_zscale} & Disable ZScale transform on image pixel values & False & N/A \\%
\emph{-{}-zscale\_contrasts} & ZScale contrast values applied to each RGB channel & 0.25,0.25,0.25 & 0-1,0-1,0-1 \\%
\emph{-{}-biascontrast} & Apply bias contrasting to image pixel values & False & N/A \\%
\emph{-{}-bias} & Bias parameter value (if \emph{-{}-biascontrast} is given) & 0.5 & 0-1 \\%
\emph{-{}-contrast} & Contrast parameter value (if \emph{-{}-biascontrast} is given) & 1.0 & 0-1 \\%
\emph{-{}-no\_norm\_img} & Disable input image normalisation & False & N/A \\%
\emph{-{}-dataloader} & What dataloader type to use & datalist & datalist, datalist\_json, \\%
& & & datadir\_json\\%
\emph{-{}-datalist} & Path to the dataset file list in the required format & \textless{}none\textgreater{} & Any path \\%
\emph{-{}-datalist\_train} & Path to the training set file list if the dataset has already been split & \textless{}none\textgreater{} & Any path \\%
\emph{-{}-datalist\_val} & Path to the validation file set if the dataset has already been split & \textless{}none\textgreater{} & Any path \\%
\emph{-{}-datadir} & Path to the top directory of the dataset & \textless{}none\textgreater{} & Any path \\%
\emph{-{}-validation\_data\_fract} & What fraction of the dataset to dedicate to the validation set & 0.1 & 0-1 \\%
\emph{-{}-maxnimgs} & The max number of images to consider in the dataset & -1 (all) & Integers \\%
\emph{-{}-class\_dict} & Class name-id dictionary to be used in dataset loading & \{spurious: 1, compact: 2,& Any \\%
& & extended: 3, & \\%
& & extended-multisland: 4, & \\%
& & flagged: 5\}& \\%
\hline%
\rowcolor{lightgray}
\multicolumn{4}{c}{TRAIN OPTIONS} \\%
\hline%
\emph{-{}-ngpu} & Number of GPUs to use & 1 & Integers \\%
\emph{-{}-nimg\_per\_gpu} & Number of images per GPU & 1 & Integers \\%
\emph{-{}-logs} & Where to store logs and weights files & logs/ & Any path \\%
\emph{-{}-nthreads} & Number of worker threads & 1 & Integers \\ %
\emph{-{}-nepochs} & Number of epochs to train the model for & 1 & Integers \\%
\emph{-{}-epoch\_length} & Number of data batches per epoch & None (=all sample) & Integers \\%
\emph{-{}-nvalidation\_steps} & Number of validation data batches per epoch & None (=all sample) & Integers \\%
\emph{-{}-classdict\_model} & Class name-id dictionary to be used in model training/testing & equal to class\_dict & Any \\%
\emph{-{}-weights} & NN weights file to use in training/testing/detection & empty & Any path to a\\%
& If empty, weights are randomly initialized & & valid .h5 file \\%
\emph{-{}-rpn\_anchor\_scales} & RPN Anchor Scales to use & 4, 8, 16, 32, 64 & Series of Integers \\%
\emph{-{}-max\_gt\_instances} & Max GT instances & 300 & Integers \\%
\emph{-{}-backbone} & Backbone network to use & resnet101 & resnet101,resnet50 \\%
\emph{-{}-backbone\_strides} & Backbone strides to use & 4, 8, 16, 32, 64 & Series of Integers \\%
\emph{-{}-rpn\_train\_anchors\_per\_image} & Number of anchors to use per image & 512 & Integers \\%
\emph{-{}-rpn\_nms\_threshold} & RPN non-maximum-suppression threshold to use & 0.7 & 0-1 \\%
\emph{-{}-rpn\_train\_anchors\_per\_image} & Number of anchors to use per image & 512 & Integers \\%
\emph{-{}-train\_rois\_per\_image} & Number of ROIs to feed to classifier per image & 512 & Integers \\%
\emph{-{}-rpn\_anchor\_ratios} & RPN Anchor Ratios to use & 0.5, 1, 2 & Series of Numbers \\%
\emph{-{}-rpn\_class\_loss\_weight} & RPN Classification Loss weight modifier & 1 & Number \\%
\emph{-{}-rpn\_bbox\_loss\_weight} & RPN Bounding Box Loss weight modifier & 1 & Number \\%
\emph{-{}-mrcnn\_class\_loss\_weight} & Mask R-CNN Classification Loss weight modifier & 1 & Number \\%
\emph{-{}-mrcnn\_bbox\_loss\_weight} & Mask R-CNN Bounding Box Loss weight modifier & 1 & Number \\%
\emph{-{}-mrcnn\_mask\_loss\_weight} & Mask R-CNN Mask Loss weight modifier & 1 & Number \\%
\emph{-{}-(no\_)rpn\_class\_loss} & Whether to use RPN Class Loss & True & N/A \\%
\emph{-{}-(no\_)rpn\_bbox\_loss} & Whether to use RPN Bounding Box Loss & True & N/A \\%
\emph{-{}-(no\_)mrcnn\_class\_loss} & Whether to use Mask R-CNN Class Loss & True & N/A \\%
\emph{-{}-(no\_)mrcnn\_bbox\_loss} & Whether to use Mask R-CNN Bounding Box Loss & True & N/A \\%
\emph{-{}-(no\_)mrcnn\_mask\_loss} & Whether to use Mask R-CNN Mask Loss & True & N/A \\%
\emph{-{}-weight\_classes} & Indicates that classes should be weighted & False & N/A \\%
\emph{-{}-exclude\_first\_layer\_weights} & Exclude first layer weights when training & False & N/A\\%
\emph{-{}-no\_augmentation} & Disable image augmentation in training & False & N/A \\%
\hline%
\rowcolor{lightgray}
\multicolumn{4}{c}{TEST OPTIONS} \\%
\hline%
\emph{-{}-remap\_classids} & Remap class ids of detected objects & False & N/A \\
\emph{-{}-classid\_remap\_dict} & Class name-id dictionary used to remap detected object & equal to class\_dict & Any \\%
& class ids (if \emph{-{}-remap\_classids} is given) & & \\%
\emph{-{}-scoreThr} & Object detection score threshold to be used during evaluation & 0.7 & 0-1 \\%
\emph{-{}-iouThr} & IOU threshold used to match detections with true objects & 0.6 & 0-1 \\%
& during testing/evaluation & & \\%
\hline%
\rowcolor{lightgray}%
\multicolumn{4}{c}{DETECT OPTIONS} \\%
\hline%
\emph{-{}-image} & Image input on which detection should be run & \textless{}none\textgreater{} & Any path to an image \\%
\emph{-{}-xmin} & From which x coordinate the image should be read & -1 (all) & Integers \\%
\emph{-{}-xmax} & Up to which x coordinate the image should be read & -1 (all) & Integers \\%
\emph{-{}-ymin} & From which y coordinate the image should be read & -1 (all) & Integers \\%
\emph{-{}-ymax} & Up to which y coordinate the image should be read & -1 (all) & Integers \\%
\emph{-{}-detect\_outfile} & Filename the generated detection plot should be stored in & \textless{}none\textgreater{} & Any filename \\%
\emph{-{}-detect\_outfile\_json} & Filename the json with detections should be stored in & \textless{}none\textgreater{} & Any filename \\%
\hline%
\rowcolor{lightgray}%
\multicolumn{4}{l}{PARALLEL PROCESSING OPTIONS} \\%
\hline%
\emph{-{}-split\_img\_in\_tiles} & Indicates that the input image should be divided into tiles & False & N/A \\%
\emph{-{}-tile\_xsize} & Size of each tile (width) & 512 & Integers \\%
\emph{-{}-tile\_ysize} & Size of each tile (height) & 512 & Integers \\%
\emph{-{}-tile\_xstep} & Tile partition grid step size along x/y, e.g. how many steps  & 1 & 0-1 \\%
\emph{-{}-tile\_ystep} & (in fraction of tile) a tile is moved wrt the previous one. & &\\%
 &  (1=no overlap, 0.5=half overlap, 0.3=70\% overlap) & &\\%
\hline%
\end{longtable}
}

\newpage%

\section{Additional figures}
\setcounter{figure}{0}
\begin{figure*}[!ht]
\centering%
\subtable[$C_{\tiny{\text{compact}}}$]{\includegraphics[scale=0.4]{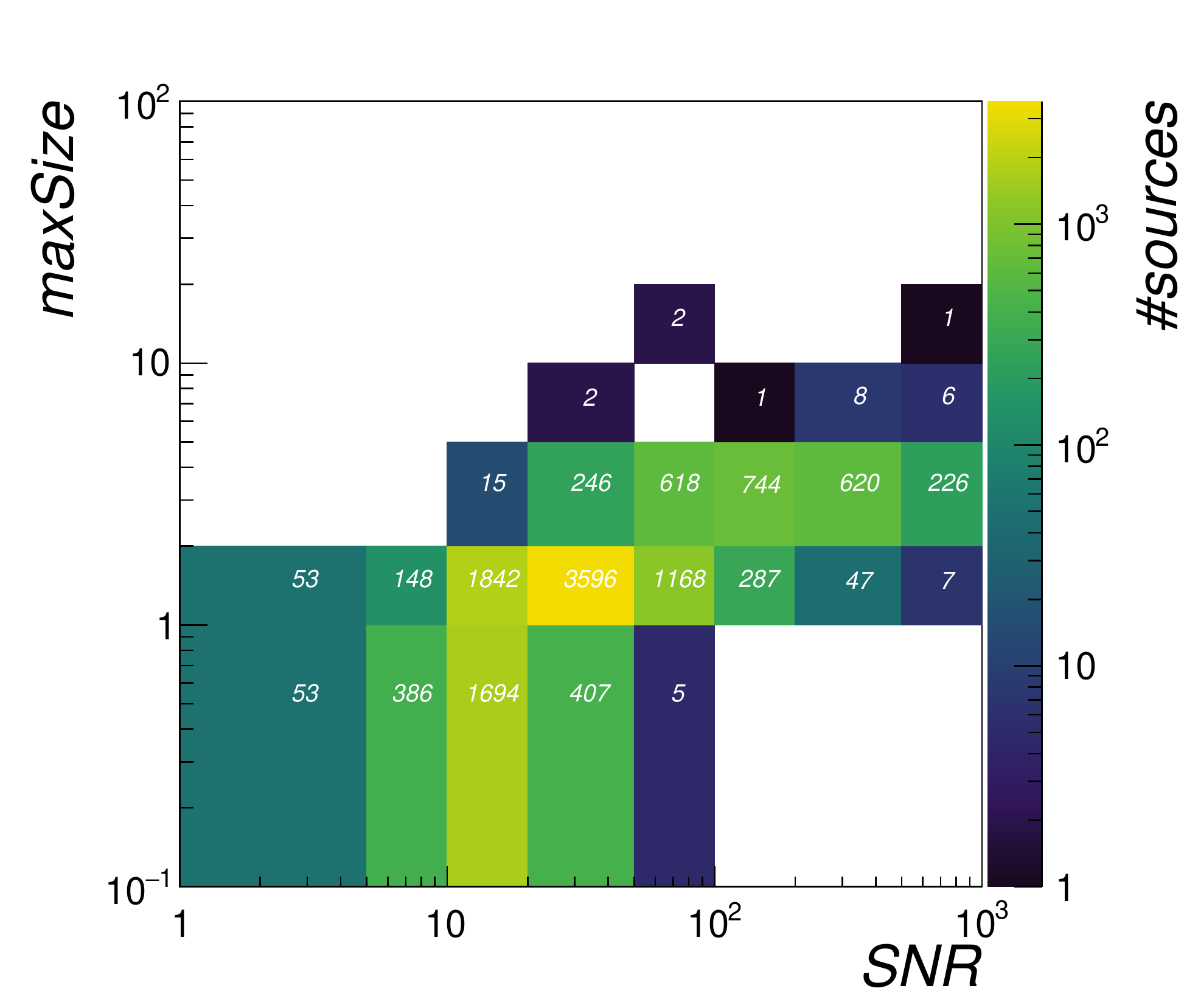}}%
\subtable[$R_{\tiny{\text{compact}}}$]{\includegraphics[scale=0.4]{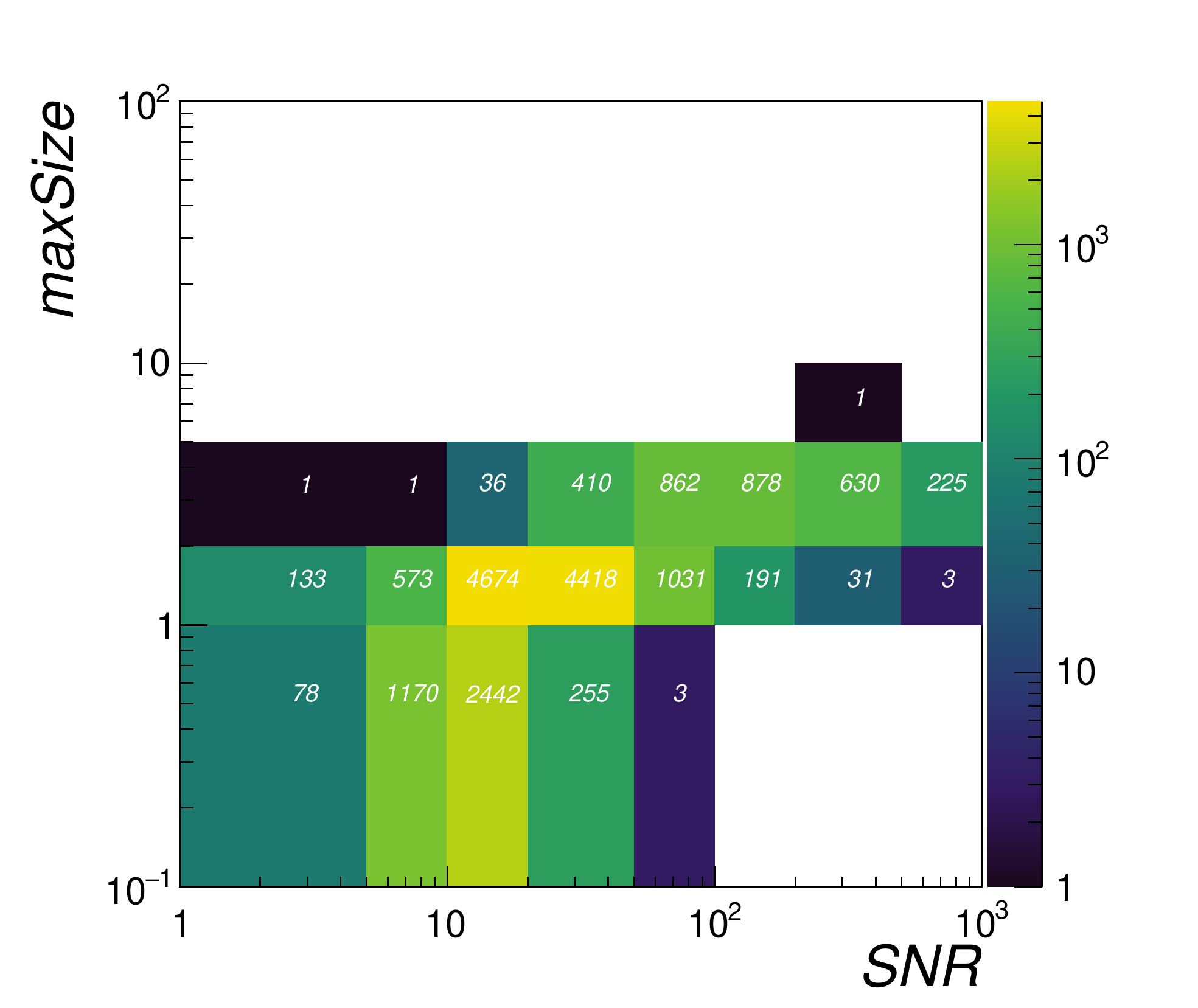}}%
\\%
\vspace{-0.42cm}
\subtable[$C_{\tiny{\text{extended+extended-multi}}}$]{\includegraphics[scale=0.4]{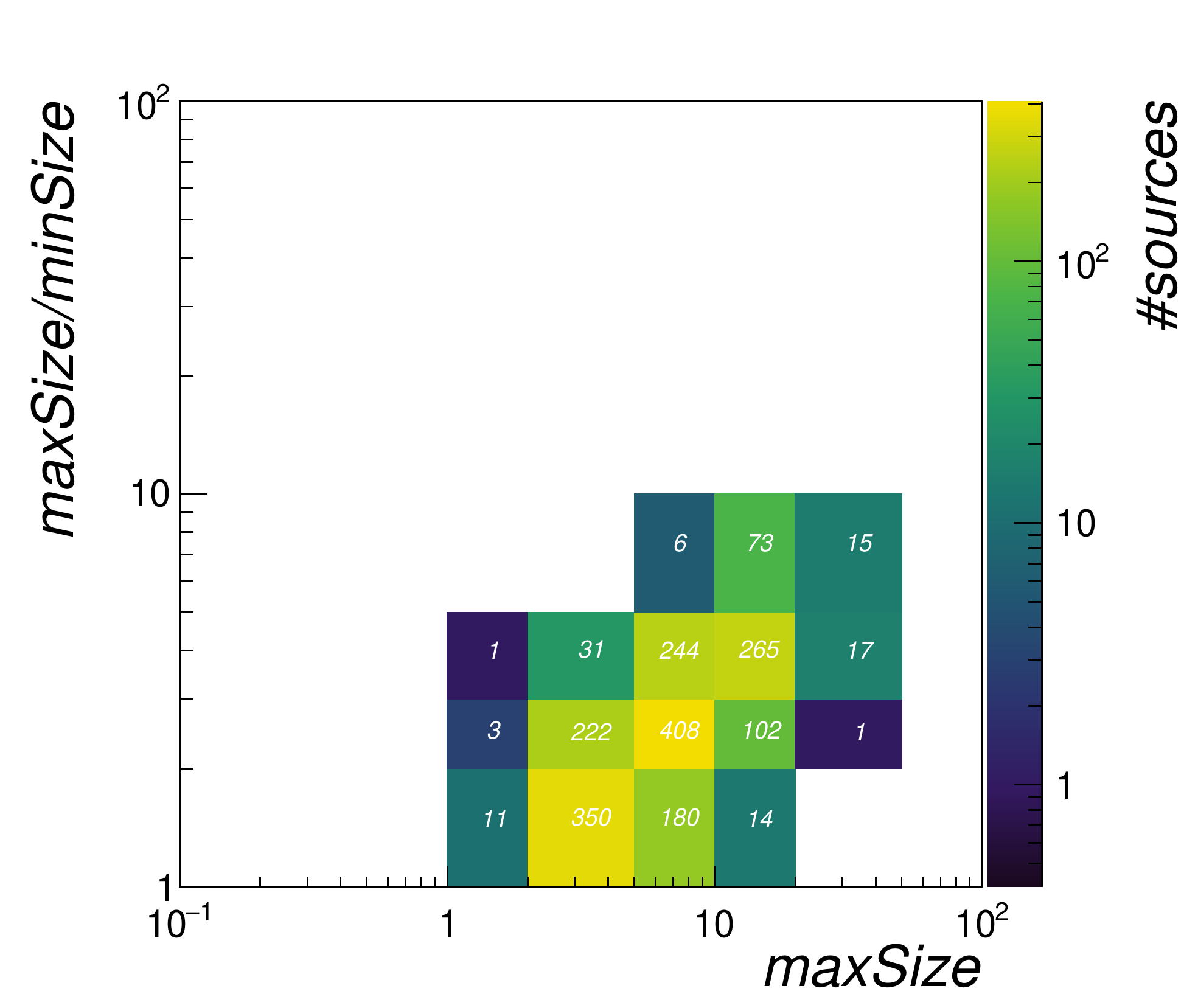}}%
\subtable[$R_{\tiny{\text{extended+extended-multi}}}$]{\includegraphics[scale=0.4]{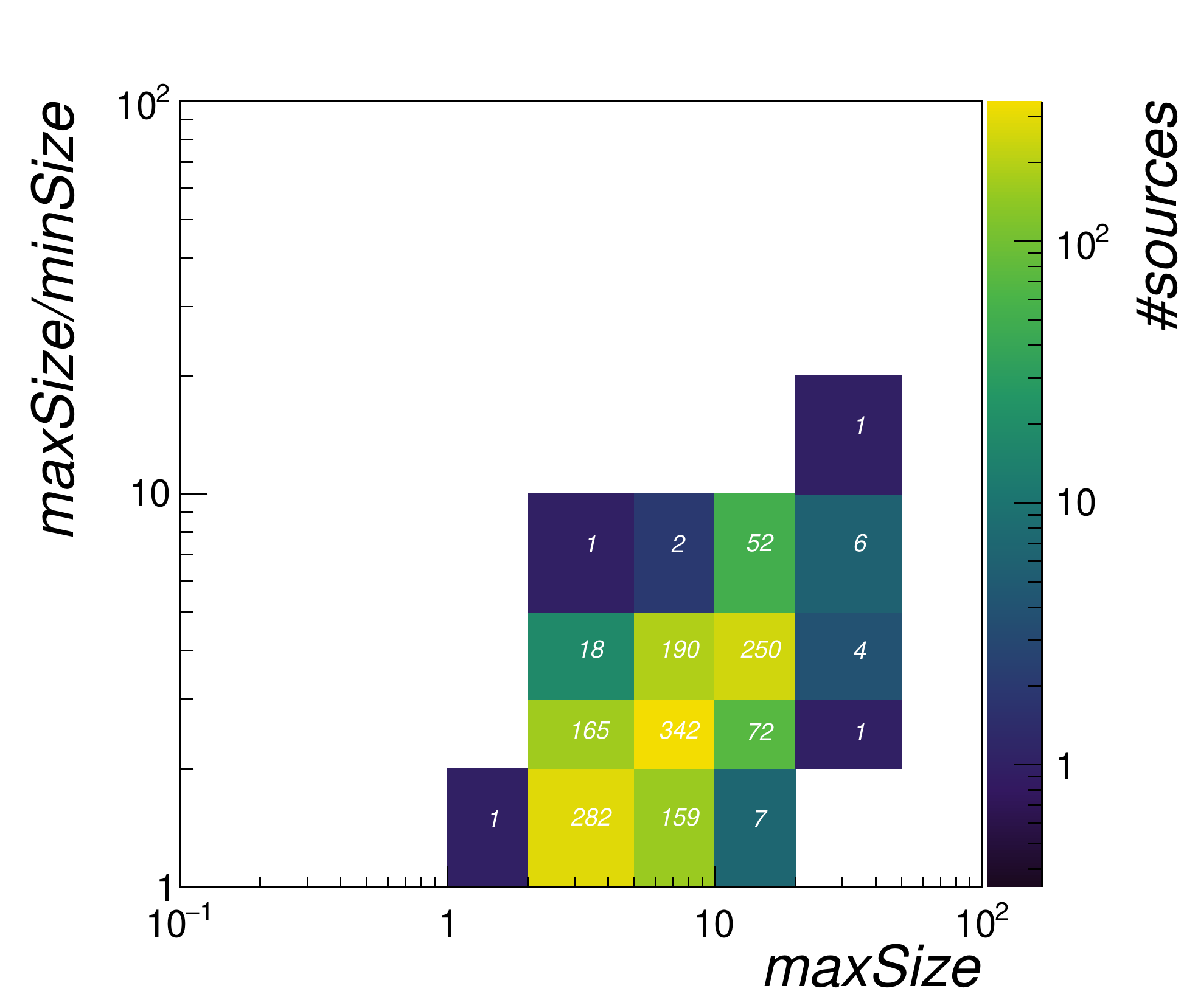}}%
\caption{%
Top: Number of sources (i.e. source counts) used to compute the completeness and reliability metrics reported in Fig.~\ref{fig:detection-perf} per each 2D bin.
}%
\label{fig:source-counts}
\end{figure*}

\end{document}